\documentclass[12pt]{article}
\usepackage{amsmath}
\usepackage{times}
\usepackage{graphicx}
\usepackage{color}
\usepackage{multirow}
\usepackage[square,sort,comma,numbers]{natbib}
\usepackage{rotating}
\usepackage{bbm}
\usepackage{latexsym}
\DeclareGraphicsExtensions{.eps,.png}
\usepackage{epstopdf}
\usepackage{epsf}
\usepackage{epsfig}
\usepackage{subcaption}
\usepackage{mathtools}

\usepackage{verbatim}

\newcommand{\captionfonts}{\normalsize}

\makeatletter  
\long\def\@makecaption#1#2{%
  \vskip\abovecaptionskip
  \sbox\@tempboxa{{\captionfonts #1: #2}}%
  \ifdim \wd\@tempboxa >\hsize
    {\captionfonts #1: #2\par}
  \else
    \hbox to\hsize{\hfil\box\@tempboxa\hfil}%
  \fi
  \vskip\belowcaptionskip}
\makeatother   

\begin{document}
\hspace{13.9cm}1

\ \vspace{20mm}\\

{\LARGE Learning Feedforward and Recurrent Deterministic Spiking Neuron Network Feedback Controllers}

\ \\
{\bf \large Tae Seung Kang$^{\displaystyle 1}$ and Arunava Banerjee$^{\displaystyle 1}$}\\
{$^{\displaystyle 1}$Computer and Information Science and Engineering Department,
	University of Florida.}\\
%


\thispagestyle{empty}
\markboth{}{NC instructions}
\ \vspace{-0mm}\\
%
\begin{center} {\bf Abstract} \end{center}

We address the problem of learning feedback control where the controller is a
network constructed solely of deterministic spiking neurons. In contrast to
previous investigations that were based on a spike rate model of the neuron,
the control signal here is determined by the precise temporal positions of
spikes generated by the output neurons of the network. We model the problem
formally as a hybrid dynamical system comprised of a closed loop between a
plant and a spiking neuron network. We derive a novel synaptic weight update
rule via which the spiking neuron network controller learns to hold process
variables at desired set points. The controller achieves its learning objective
based solely on access to the plant's process variables and their derivatives
with respect to changing control signals; in particular, it requires no
internal model of the plant. We demonstrate the efficacy of the rule by
applying it to the classical control problem of the cart-pole (inverted
pendulum) and a model of fish locomotion. Experiments show that the proposed
controller has a stability region comparable to a traditional PID controller,
its trajectories differ qualitatively from those of a PID controller, and in
many instances the controller achieves its objective using very sparse spike
train outputs.

\section{Introduction}

Animals are exquisite control systems. Whether it be the flight of a dragonfly
or the walking of a biped, state-of-the-art engineered systems are yet to
achieve the versatility and robustness displayed by their animal counterparts.
In addition, in many instances the particular control task, locomotion for
instance, is {\em learned} by the animal. Our goal in this work is to address
this question of learning to control in the context of a biologically motivated
constraint---the fact that the vast majority of neurons in animal brains
communicate with one another using action potentials, also known as spikes.

In higher animals, several neural sub-systems interact synergistically to
achieve the overall control objective. In vertebrates, for example, the control
signals received by the effector skeletal muscle fibres are in the form of
spike trains generated by lower motor neurons \citep{squire2012fundamental}.
The controller itself is a network of spiking neurons that resides upstream
from the lower motor neurons, hypothesized to be located in the basal ganglia
and the cerebellum \citep{squire2012fundamental}. The controller receives
inputs, which in the case of a feedback controller are process variables that
are to be maintained at fixed or dynamically varying set points. The process
variable input into the controller is in turn computed elsewhere and
incorporates the combined and integrated output of one or more sensory systems
as well as the output of the muscle spindles.


To formally delineate the problem of learning a spiking neuron network
controller, we abstract away all aspects of the system that are of secondary
concern and replace them with fixed predefined modules. In particular, we model
the entire process beginning at the spike train output of the controller and
culminating at the control signal generated (such as the force exerted by the
muscle fibres) using fixed functions of the spike train. Although our framework
can accommodate any deterministic, differentiable mapping from a bounded time
window of the spike train output of the controller to a continuous time control
signal, for the sake of clarity, we focus on functions that are additively
separable. In effect, the continuous time control signal is generated by
convolving the spike train output of the controller with fixed causal
differentiable kernels of bounded support.

The impact of the control signal on the organism immersed in its environment,
we model using a fixed plant. Finally, we model the input of the process
variables and their dynamically varying set points as postsynaptic potential
inputs into specifically identified neurons of the controller. Our objective is
to devise a formal synaptic weight update rule that when applied to the neurons
of the controller, induces the controller to learn to perform the control task.
We have confined ourselves to a framework where the controller is allowed
access solely to derivatives of the plant's process variables with respect to
changing control signals, to achieve its learning objective. In particular,
the controller does not have access to any internal forward model of the plant.

That the above problem differs from those previously studied in feedback
control, can be discerned from the following observation. Traditional feedback
controllers such as the proportional-integral-derivative (PID) controller or
its variants are designed to solve a control problem in the continuous domain
with few restrictions. The process variable is a bounded continuous function of
time, and so is the control signal generated by the controller; there is little
else that constrains these functions. In contrast, the control output generated
by the spiking neuron network controller is an ensemble of spike trains. The
spike trains when convolved with the fixed convolution kernels referred to
above, lead to a highly restricted and stereotyped signal. In particular, it
is easy to observe that given a kernel, there exists a bound $C$ such that any
non-zero control signal $u$ satisfies $|\!|u|\!|_\infty>C$---informally, the
controller has the choice between generating no output or an output larger than
a fixed amplitude. This has immediate implications for the stability of the
fixed point (determined by the set point) of the combined (the controller and
the plant in closed loop) dynamical system; the process variable can at best be
made to oscillate around the set point.

Finally, we emphasize that the present work considers a controller that lacks
an explicit internal model \citep{francis1976internal} of the plant. How an
internal model may represent the putative future state of the plant using spike
trains, and how such an internal model may be learned, are complex problems in
their own right that are outside the scope of this article.


The remainder of the paper is structured as follows. Following a review of
related work in Section~\ref{review}, we formally define the coupled dynamical
system framework in Section~\ref{bigpicture}. We highlight the difficulty
facing analysis when operating with both continuous time signals as well as
spike times in the same model, and delineate the approach that resolves this
issue. Section~\ref{srm} then describes the neuron model, and
Section~\ref{plant} briefly describes the two plants investigated in this work,
noting their process variables and corresponding set points.
Section~\ref{feedback-control} comprises the core of our contribution where the
synaptic weight update rule is derived. Sections~\ref{simulated},
\ref{pendulum}, and \ref{fish} present experimental results from several
variations of the controller, and Section~\ref{conclusion} concludes with final
remarks.

\section{Background} \label{review}

Animal motor control investigated through the lens of control theory has a long
and rich history. Early theories aiming to explain why coordinated movement in
animals was stereotypical in spite of the existence of redundant biomechanical
degrees of freedom, appealed to inherent constraints imposed by the nervous
system as well as synergies in muscle groups \citep{tresch1999construction, d2005shared, d2006control, ting2005limited}. These theories have since been supplanted by what is now the dominant viewpoint in the field
advanced by optimal control theory. This view posits that coordinated movement
and trajectory planning can be formally posed as an optimization problem on a
cost function that not only accounts for the final goal of the intended
activity, but also penalizes path integrated considerations such as total
squared jerk \citep{flash1985coordination, hoff1993models, uno1989formation} or sum of squared motor commands \citep{zajac1989muscle}. A variation of optimal control theory, optimal feedback control theory \citep{todorov2002optimal}, incorporates instantaneous sensory feedback into this framework.
The primary goal of these theories is to elucidate the nature of the control
policies and they, therefore, do not address how such control policies may have
come to be implemented in neuronal hardware.





In a largely independent strand of research, substantial strides have been made
in recent years in learning in feedforward networks of spiking neurons. One of
the early results was that of the {\em SpikeProp} supervised learning rule
\citep{bohte2002error} where a feedforward network of spiking neurons was
trained to generate a desired pattern of spikes in the output neurons in
response to an input spike pattern of bounded length, with the caveat that each
output neuron spike exactly once in the prescribed time window during which the
network received the input. Posing the problem differently, \cite{tempotron}
proposed the {\em Tempotron} that learned to discriminate between two sets of
bounded length input spike trains by generating an output spike in the first
case and remaining quiescent in the second. Subsequent results
\citep{florian2012chronotron, mohemmed2012span,memmesheimer2014learning, Banerjee:2016:LPS:2979100.2979103} have
relaxed most constraints to the extent that one can now learn precise spike
train to spike train transformations in feedforward networks of spiking neurons
via a general synaptic weight update rule.



 
Thus far, attempts to model controllers using spiking neuron networks have used
a spike rate based coding of the underlying continuous time signals. For
example, \cite{bouganis2010training} have proposed a spiking neuron network
based controller that learns using spike time dependent plasticity
\citep{song2000competitive}. The controller displays high firing rates due to
the underlying rate based code. Elsewhere, genetic algorithms have been used to
construct such controllers, again within the rate code paradigm
\citep{floreano2001evolution,batllori2011evolving, hagras2004evolving, wiklendt2014locomotion,takase2015evolving}. Finally,
\citep{fremaux2013reinforcement,hennequin2014optimal} have studied control
problems under the framework of reinforcement learning, where the actor-critic
model \citep{barto1983neuronlike,anderson1987strategy} is used to train the
networks.




The synaptic weight update rule that is derived in this article generalizes the
model presented in \cite{Banerjee:2016:LPS:2979100.2979103} to recurrent
networks and, furthermore, incorporates into the framework neurons that not
only receive spike input but also continuous time postsynaptic potential inputs
representing the process variables and their dynamically varying set points.
The feedback control problem is then turned into a learning problem where the
objective is for the controller to learn synaptic weights so as to be able to
hold process variables at prescribed set points. Theoretical and experimental
results restricted to feedforward networks were presented in
\cite{kang2017learning}.

%
%

%

\section{Framework} \label{bigpicture}

\begin{figure}
	\centering
	\begin{subfigure}{0.6\textwidth}
		\includegraphics[width=\textwidth]{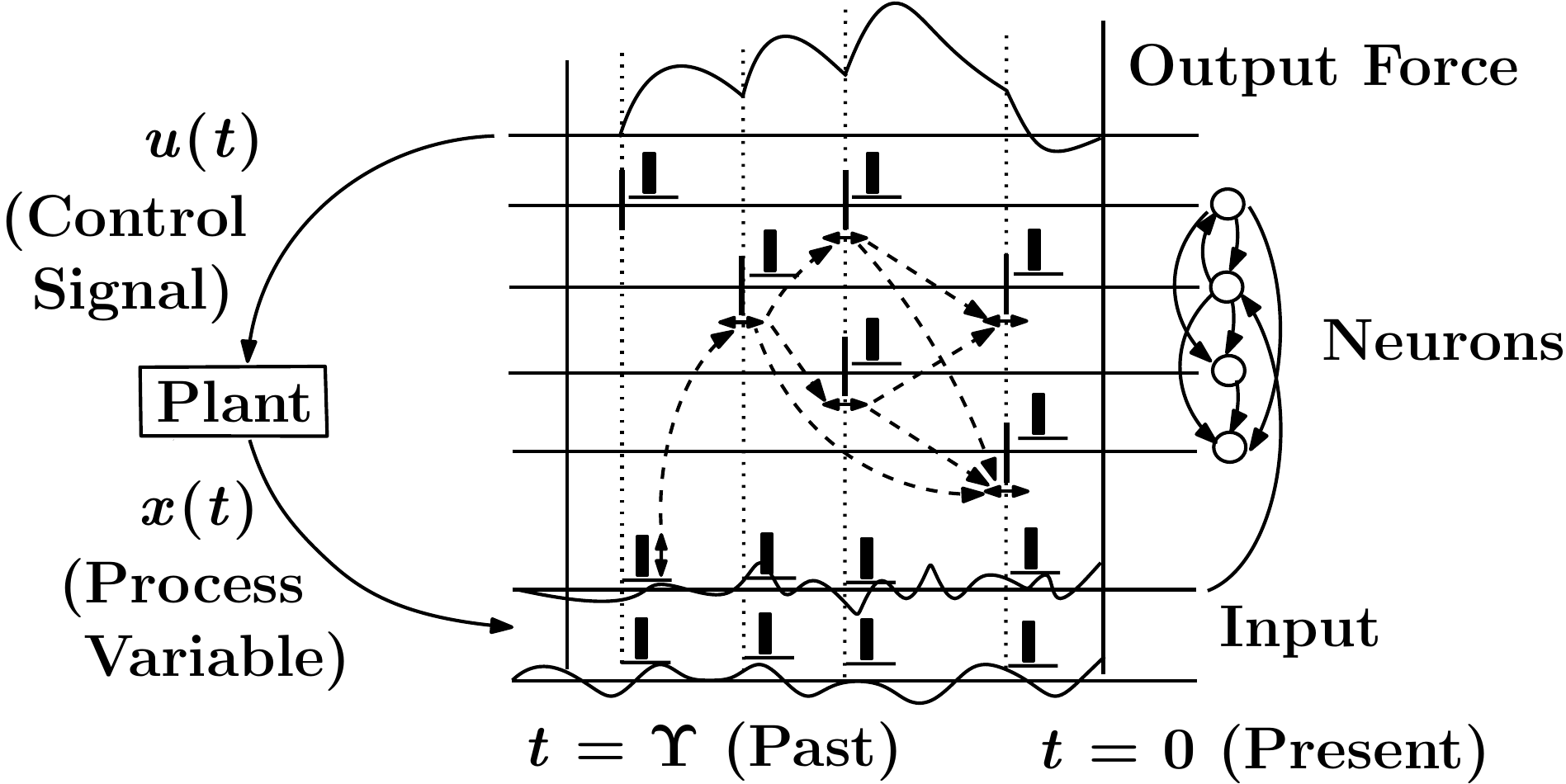}
		\caption{Plant and Controller}
		\label{a}
	\end{subfigure}
	\hspace{1em}
	\begin{subfigure}{0.34\textwidth}
		\includegraphics[width=\textwidth]{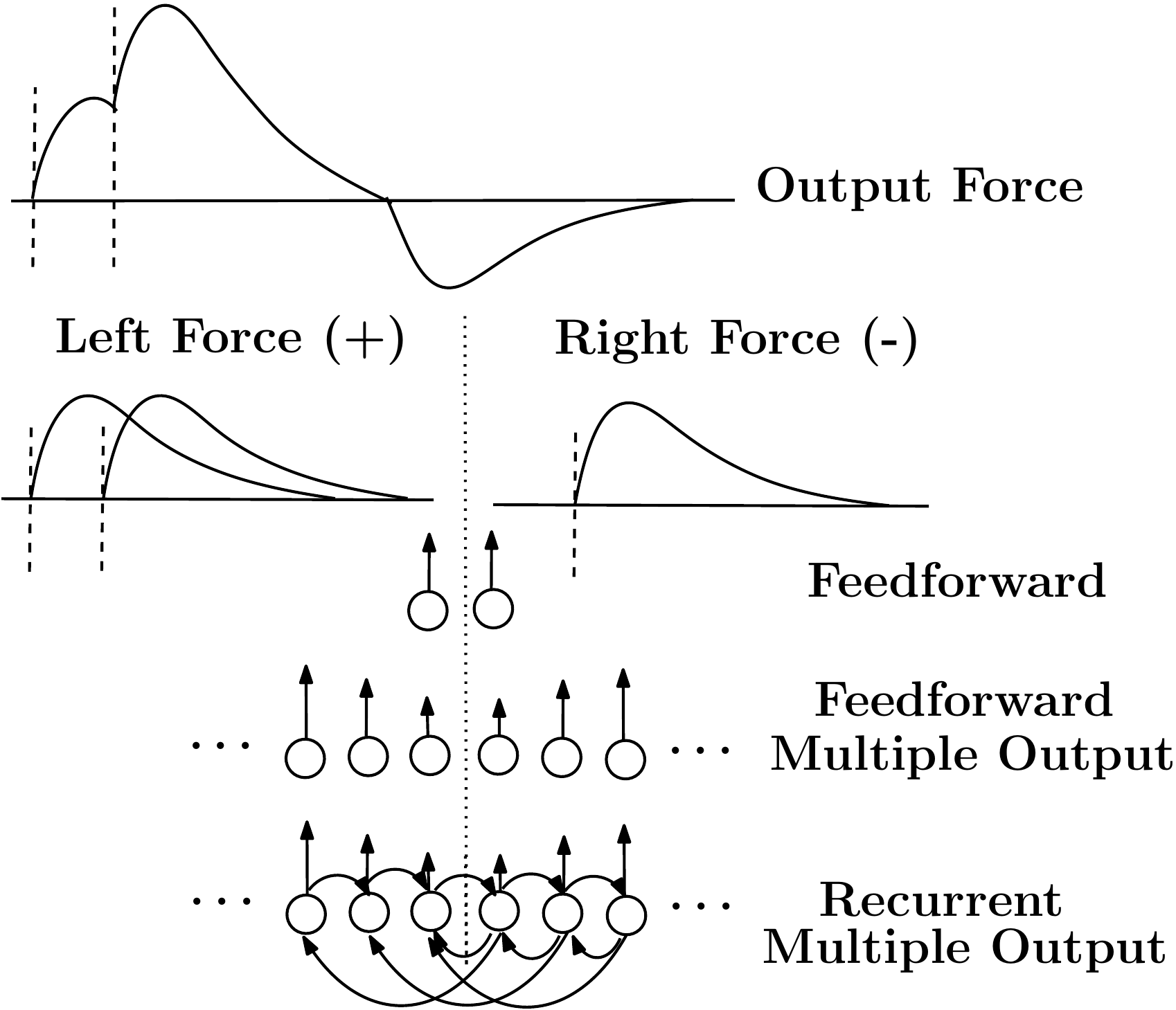}
		\caption{Models}
		\label{b}
	\end{subfigure}
	\caption{
		(a)	The hybrid dynamical system that models the control problem. The
		plant receives control input $u(t)$ and outputs process variable $x(t)$
		according to $\dot x=f(x,u,t)$. The controller is a feedforward or
		recurrent network of spiking neurons that takes as input $x(t)$ and
		generates output spike trains, which are then convolved with fixed
		kernels to produce $u(t)$. 
		Horizontal lines with arrows represent perturbations in spike times.
		Vertical bars represent weights assigned to spikes, or in the case of
		continuous time signals, to weights assigned to temporal segments of
		the signal. Dotted arrows indicate how perturbations in weights
		perturb spike times of future spikes generated in the network.
		Perturbing the synaptic weights of any neurons in the network, in the
		past, perturbs the output spike times of the output neurons in the
		network, which when convolved with the kernel creates a perturbation in
		the control signal. The controller is optimized based on an error
		function that embodies the deviation of the process signal from its set
		point.
		(b) Models. We consider three models: Feedforward network with single
		kernel, Feedforward network with multiple kernels, and recurrent
		network with multiple kernels.
	}
	\label{fig:overview}
\end{figure}

Our objective is described schematically in Figure~\ref{fig:overview}. We
consider a hybrid dynamical system comprised of a closed loop between a plant
and a network of spiking neurons that models the controller. The plant is, in
general, governed by a non-linear differential equation, $\dot x= f(x,u,t)$
where $x(t)$ and $u(t)$ are the vector valued process and control variables,
respectively. We consider the scenario where the controller does not have an
explicit internal model of $f(\cdot)$. The objective of the controller is
merely to hold the process variable $x(t)$ as close as possible to the set
point $x^*(t)$ by generating a control signal $u(t)$. The set point $x^*(t)$
may have been computed by a forward model present elsewhere in the system.
Furthermore, our framework is agnostic to whether the trajectory $x^*(t)$ is
precomputed, or is computed on-line based on the current state of the plant.

All neurons in the controller network generate spikes as output,
deterministically, according to a minor variation of a standard neuron model
that we describe in Section~\ref{srm}. The input neurons of the controller
receive the continuous time $x(t)$ (and $x^*(t)$ in the case of a time varying
set point) as postsynaptic potentials in addition to postsynaptic potentials
induced by spikes received from other neurons in the network. That the behavior
of a putative controller would be different from a standard feedback controller
follows immediately from the following observation. Assuming finite non-zero
thresholds for all neurons in the network and a simple scenario of $x^*(t)=0$,
it is clear that the input neurons will not generate any spikes for
sufficiently small $x(t)$, and therefore the controller cannot react to
deviations away from the set point that are below a sensitivity threshold.

Each output neuron of the controller generates spike trains that are then
mapped to a control signal $u(t)$ via a deterministic function
$\kappa(t^O_1,t^O_2, \ldots)$, where $t^O_i$ is the $i$th output spike time of
the output neuron. By convention, $t^O_i$ is a positive real number denoting
the time that has elapsed since the generation of spike $i$. Although our
framework seamlessly generalizes to any differentiable function with bounded
support, for the sake of clarity we consider an additively separable
differentiable function with bounded support, that is,
$u(t)=\sum_i\kappa(t^O_i)$ where $\kappa(t)=0$ for $t<0$ and $t>\Upsilon$. If
the spikes were to be modeled as Dirac delta functions, this could
alternatively be viewed as convolving the spike train with a fixed causal
kernel of bounded support. Once again the difference in behavior of a putative
controller is clear from the observation that not only is $u(t)$ highly
stereotyped, but there also exists a bound $C$ such that any non-zero control
signal $u$ satisfies $|\!|u|\!|_\infty>C$.

The controller network is parameterized by the set of synaptic weights of all
synapses in the network. Our overall goal is to demonstrate that these synaptic
weights can be \emph{learned}. Our approach is based on identifying whether
small perturbations in the synaptic weights in the past could have led to a
slightly superior control signal at present. Superior is objectivized by
determining whether the process variable $x(t)$ would then have been closer to
the control point $x^*(t)$ at the current time $t=0$. The network is then
trained using stochastic gradient descent on this objective. Since as noted
earlier, the network has an intrinsic sensitivity threshold, the stopping
criterion for learning is based on this threshold.

The primary difficulty in the problem arises from the need to model continuous
time signals such as $x(t)$ or $u(t)$ and spike times of the neurons in the
network, under a common framework. To resolve this problem we negotiate several
refinements. First, we approach the problem via a perturbation analysis.
Whereas the continuous time signals are perturbed in the range, as is standard,
spikes, on the other hand, are set as stereotyped objects that can only be
perturbed in time (indicated by horizontal arrows in
Figure~\ref{fig:overview}a). Second, we assume that the neurons in the network
are bounded memory devices; the effect of a past spike on the current membrane
potential of a neuron is 0 if the spike has aged beyond a time bound
$\Upsilon$. Noting that neurons have an absolute refractory period that
prevents successive spikes from occurring closer than a certain time bound, we
can conclude that there can only be finitely many spikes in the window
$[0,\Upsilon]$ that have an impact on the current membrane potentials of the
neurons in the network. The perturbation analysis, consequently, is confined to
finitely many spikes in the past. Third, noting that any learning algorithm
updates synaptic weights, successive spikes arriving at the same synapse can
have different effects on the postsynaptic potential of the neuron. We
accommodate this in the analysis by virtually assigning synaptic weights to
spikes rather than to the synapse (indicated by vertical bars in
Figure~\ref{fig:overview}a). Likewise, for the continuous time signals, we
assign synaptic weights to segments of the signal in the past. Fourth, we
restrict synaptic weight updates to those instants in time where spikes are
generated by the output neurons of the network. Finally, noting that the
effects of spikes as well as of the continuous signals are causal on other
spikes generated in the network (indicated by the dotted arrows in
Figure~\ref{fig:overview}a), we surmise that regardless of the network
architecture, whether it be feedforward or recurrent, the effect of
perturbations form a partial order in time. Brought together, these refinements
result in a well defined synaptic weight update rule as is demonstrated in
upcoming sections.

We probe three model architectures of successively increasing complexity, based
on connectivity and the number of output neurons in the controller network and
their corresponding $\kappa(\cdot)$ functions (Figure~\ref{fig:overview}b).
Model 1, the simplest model, is defined as a network with the simplest possible
architecture, that is no hidden layers. In order to model a $u(t)$ whose value
can be positive or negative, we use two output neurons with positive valued
$\kappa(\cdot)$ and set $u(t)$ to be their difference. Model 2, the model of
intermediate complexity, is defined as a network with four or more output
neurons (groups of two or more) with distinct $\kappa(\cdot)$'s, with a
feedforward architecture comprised of one or more hidden layers. Model 3, the
most general model, is defined as a recurrent network where the output neurons
are fully connected with each other as well as other neurons in the network.
The synaptic weight update rule that we derive is general and applies to all
architectures.


\section{Neuron Model} \label{srm}

We use a minor variation of the Spike Response Model
\citep{gerstner2002spiking} for the neurons in our controller. The membrane
potential of a neuron is computed as the sum of postsynaptic potentials
elicited by spikes arriving at its various synapses from other neurons in the
network, and afterhyperpolarization potentials elicited by the spikes generated
by the neuron itself. A special class of neurons, the input neurons, have
process variables directly injected as additional postsynaptic potentials. We
assume in addition that all afferent (incoming) and efferent (outgoing) spikes
that were generated earlier than $t=\Upsilon$ in the past have no effect on the
present membrane potential of the neuron (See Figure~\ref{fig:overview}a). The
neuron generates a spike when the membrane potential crosses the threshold
$\Theta$ from below.

The postsynaptic potential elicited by a spike is computed as the product of
the synaptic weight at the time of the arrival of the spike at the synapse with
the prototypical postsynaptic response function assigned to that synapse.
Likewise, the postsynaptic potential elicited by the continuous time process
variable is computed as the product of the synaptic weight at the current time
with the current value of the process variable. Since the learning algorithm
updates all synaptic weights whenever spikes are generated by the output
neurons of the controller, to enable the analysis of the system over any finite
stretch of time, past synaptic weights are virtually assigned to the
corresponding spikes in the case of synapses that receive spikes, and to the
corresponding intervals of time in the case of synapses receiving continuous
time process variable inputs. Finally, each spike generated by the neuron
elicits a prototypical afterhyperpolarization potential. The present is set at
$t=0$ with $t>0$ denoting the past.

Formally, the membrane potential, $P$ of a neuron at the present time $t=0$ is
given by

\begin{align} \label{eq:srm_rnn} 
P=
\sum_{i \in \Gamma_c} \tilde{w}_{i,0} x_i (0)
+ \sum_{i \in \Gamma_s} \sum_{j \in \mathcal{F}_i} w_{i,j}~ \xi (t^I_{i,j}) 
+ \sum_{k \in \mathcal{F}} \eta ( t^O_k).
\end{align}

where $\Gamma_c$ is the set of synapses receiving the continuous time process
variable inputs, $\Gamma_s$ is the set of synapses receiving afferent spikes
from other neurons in the network, $\mathcal{F}_i$ is the set of previous
spikes that have arrived at synapse $i$, and $\mathcal{F}$ is the set of
previous spikes generated by the neuron. $\tilde{w}_{i,0}$ is the weight of
continuous time input synapse $i$ immediately after the most recent output
spike of the system, and therefore in general, $\tilde{w}_{i,l}$ is the
synaptic weight over the interval between the $l$th and $(l+1)$th most recent
output spike of the system. Since synaptic weights are updated at the times of
generation of spikes at the output neurons of the controller, it follows that
$\tilde{w}_{i,0}$ is the current weight of synapse $i$. $w_{i,j}$ is the weight
of the $j$th most recent afferent spike at synapse $i$ receiving spike input.
$\xi(\cdot)$ is the prototypical postsynaptic potential elicited by an afferent
spike and $\eta(\cdot)$ is the prototypical afterhyperpolarization potential
elicited by an efferent spike. $t^I_{i,j}$ is the time elapsed since the
arrival of the $j$th most recent spike at synapse $i$, and $t^O_k$ is the time
since the generation of the $k$th most recent efferent spike.

The functional form of $\xi(\cdot)$ that we have used (and this can be
modified without affecting the analysis) is

\begin{align} \label{eq:xi} 
\xi(t) = \frac{e^{-\beta d^2 / t}}{d\sqrt{t}}e^{-t/\tau} 
\mbox{\ \ \ for\ \ } 0<t\le\Upsilon \mbox{\ \ \ and 0 otherwise}
\end{align}

where $\beta$ and $\tau$ control the rate of rise and fall of the postsynaptic
potential, and $d$ denotes the distance of the synapse from the soma.
Likewise, the functional form of $\eta(\cdot)$ is

\begin{equation}
\eta (t) = R e^{-t/\gamma} \mbox{\ \ \ for\ \ } 0<t\le\Upsilon \mbox{\ \ \ and 0 otherwise}
\end{equation}
where $R$ denotes the instantaneous fall in potential after a spike and
$\gamma$ controls its rate of recovery.

\section{Overview of Plants} \label{plant}

\subsection{Inverted Pendulum} 

\begin{figure}
	\centering
	\begin{subfigure}{0.4\textwidth}
		\includegraphics[width=\textwidth]{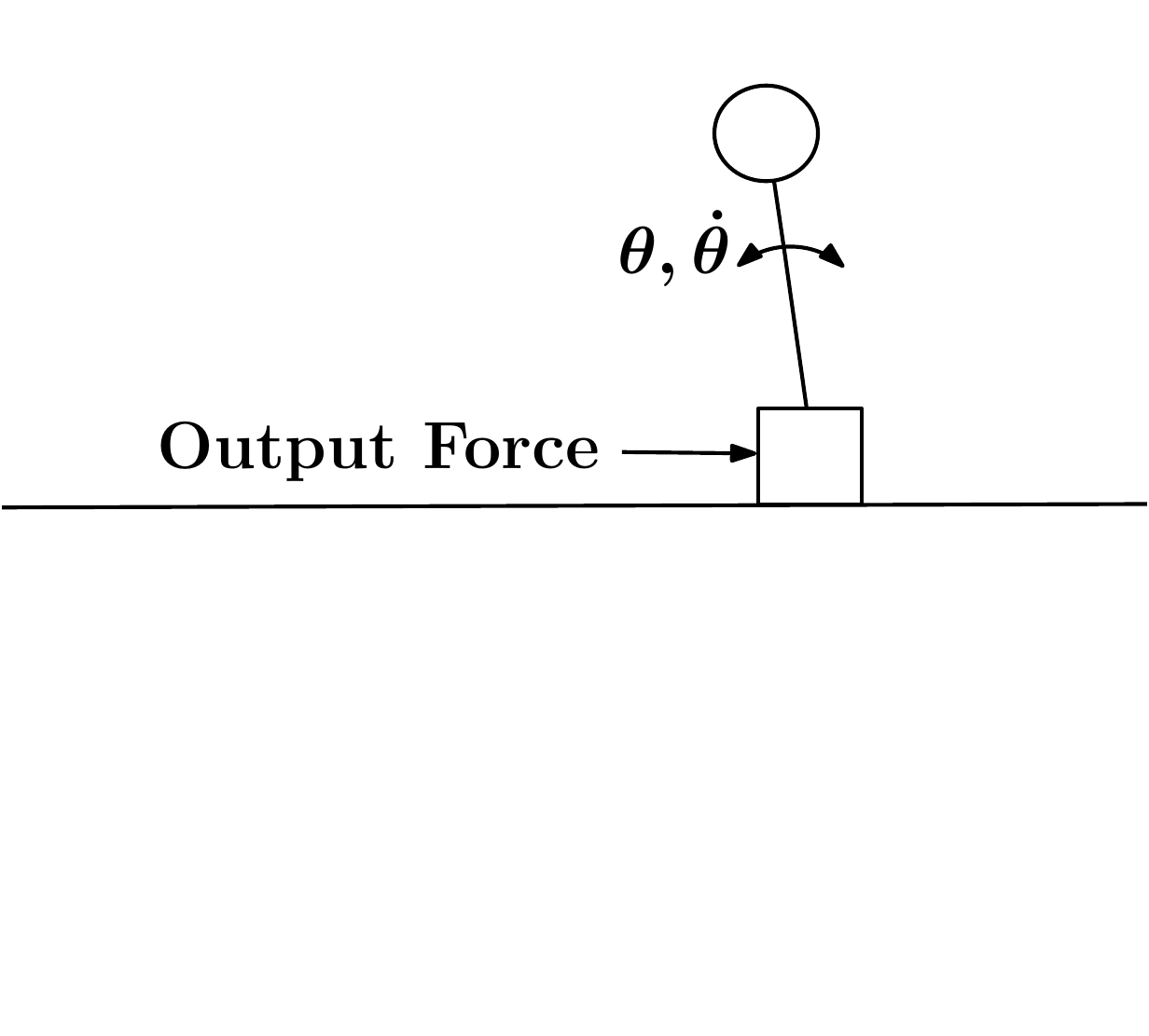}
		\caption{Inverted Pendulum}
		\label{a}
	\end{subfigure}
	\hspace{1em}
	\begin{subfigure}{0.4\textwidth}
		\includegraphics[width=\textwidth]{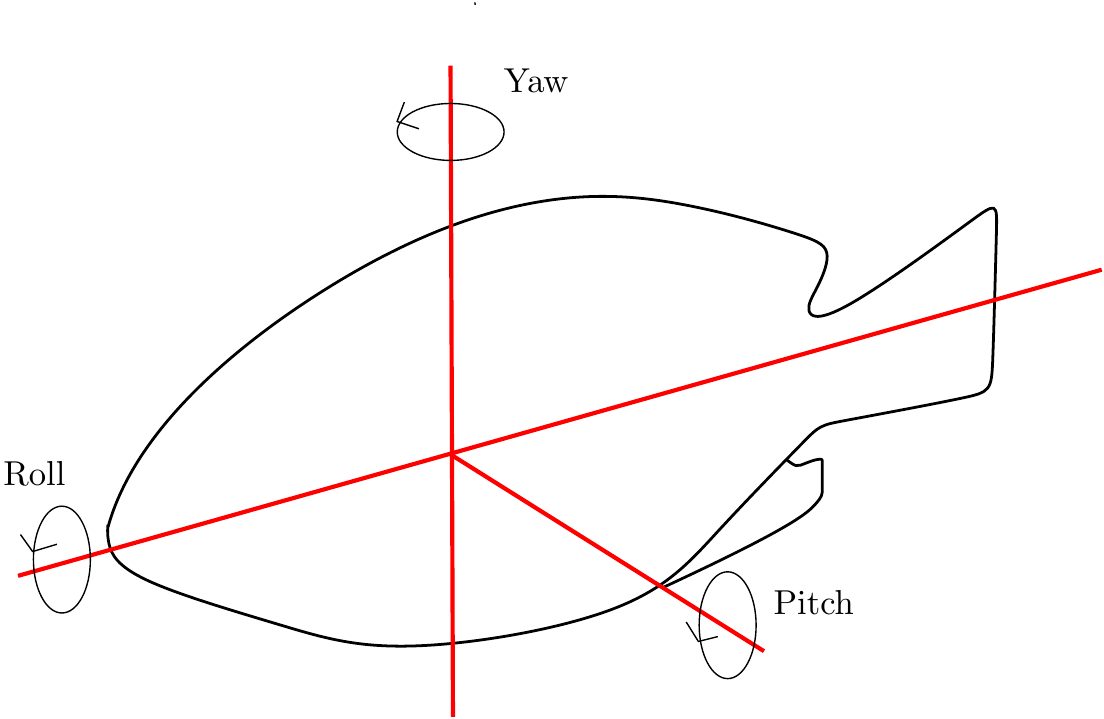}
		\caption{Fish in 3-dimensional space}
		\label{b}
	\end{subfigure}
	\caption{
		(a)	Inverted Pendulum and (b) Fish in 3-dimensional space.
	}
	\label{fig:plant}
\end{figure}

The first plant we consider in this article is the classical control problem of
the cart-pole (also known as the inverted pendulum) as shown in
Figure~\ref{fig:plant}a. The cart-pole comprises of an inverted rigid pendulum,
with the mass at the top. The pendulum is fulcrumed at its base to the cart
which rests on a frictionless surface. Control signals mapped to forces can be
applied to the cart to move it along the horizontal axis. The control problem
is to apply forces to the cart to maintain the upright position of the
pendulum. The process variables that we have considered for this plant are:
$\theta$, the angular deviation of the pendulum from the upright position, and
$\dot\theta$, the angular velocity of the pendulum. The set points for the
process variables are $\theta=0, \dot\theta=0$. The details of the system
dynamics can be found in \citep{anderson1987strategy}. All quantities of
interest as presented in Section~\ref{feedback-control}, we have derived
through numerical computations.

\subsection{Fish Locomotion} 

We also consider a more complex plant where a fish swims in 3-dimensional space
$(x, y, z)$ as depicted in Figure~\ref{fig:plant}b. The trajectory of the
fish's center of gravity ({\em cog}) can be controlled by regulating three time
varying control variables, the {\em yaw}, {\em pitch}, and {\em roll}.  This
control problem is substantially more complex since the control variables have
to be regulated in a ``synergistic'' fashion---the yaw, pitch and roll have to
interact to achieve even the simplest of locomotion tasks.  The process
variables consist of the 3-dimensional coordinate $(x, y, z)$ of the fish's
{\em cog} and its orientation $(\theta_x, \theta_y, \theta_z)$ with respect to
a target location. The orientation angles range between $-180$ and $180$
degrees with $(\theta_x, \theta_y, \theta_z)=(0,0,0)$ corresponding to the fish
facing the target.  The target location is given by $(x^*, y^*, z^*)$. We have
considered a scenario where the set point is fixed at a different
$(x^*,y^*,z^*)$ in each experiment.  The fish begins at the origin $(0, 0, 0)$
and is required to arrive at the target $(x^*,y^*,z^*)$ using a fixed velocity
profile that ramps up from 0, stays constant, and then ramps down to 0.

We have considered several variations of the problem where one or more of the
control variables are shut off and the target $(x^*,y^*,z^*)$ is set such that
the task is achievable using the smaller set of controls.  The details of the
system dynamics can be found in the SOEIL project \citep{soleil}.  Once again,
all quantities of interest as presented in Section~\ref{feedback-control}, we
have derived through numerical computations.

\section{Feedback Control using Spiking Neuron Networks}\label{feedback-control}

We now derive a synaptic weight update rule that applies to any connectivity
architecture, be it recurrent or feedforward. The basic insight is to focus on
the inputs to the neurons, spikes or continuous time, rather than the neurons
themselves, and subsequently to realize that regardless of the architecture of
the network, the spikes and continuous time inputs arrange themselves in a {\em
causal} partial order (the dotted arrows in Figure~\ref{fig:overview}a). In
essence, perturbations to the timing or the weight of a spike, or to the weight
assigned to a segment of the continuous time input, cause perturbations in the
timing of spikes that are generated at a later time, and therefore, there are
no cycles that can invalidate the application of the chain rule.

The graph structure of the impact of perturbations can, however, be very
complex, with a spike perturbation causing a perturbation in another spike via
various intersecting causal paths. Rather than enumerate all such paths, we
resolve this using a recursive/dynamic programming approach, where all effects
are computed and stored at the time of generation of each spike in the network.
Not only does this approach circumvent the issue of identifying the potentially
exponentially many paths, it also fits well with the online nature of the
updates.

\subsection{The error function and control output}

The proposed spiking neuron network controller depicted in
Figure~\ref{fig:overview} can be formally modeled as follows. Consider a plant
with the current value of the process variables represented by vector $\langle
x_1(0)$, $x_2(0)$, ..., $x_\mathcal{D}(0)\rangle$, where $\mathcal{D}$ is the
number of process variables to be controlled. The desired state of the plant
(i.e., the set point of the process variables) is represented by $\langle
x_1^*(0)$, $x_2^*(0)$, ..., $x_\mathcal{D}^*(0)\rangle$. The instantaneous
error $E$ can then be defined as $\frac{1}{2} \sum_{i=1}^{\mathcal{D}} (x_i (0)
- x_i^*(0))^2 $.  The synaptic update rule that we derive next is based on
minimizing this objective using stochastic gradient descent.

A traditional controller receives continuous time process variables from the
plant and generates a continuous time control output. The proposed controller,
however, generates spike trains, one for each neuron, instead of a continuous
output. The spike train generated by each output neuron $j$ is convolved with
the kernel 
\begin{equation}
\kappa (t) = t e^{-t / \tau_f}
\mbox{\ \ \ for\ \ } 0<t\le\Upsilon \mbox{\ \ \ and 0 otherwise}
\end{equation}
to generate a continuous time control output (which we shall henceforth refer
to as force due to the nature of the plants we have considered in this
article):
\begin{equation}
F_j = \mu_j \sum_{i \in \mathcal{F}_j} \kappa ( {^j t_i^O})
\end{equation}
where $\tau_f$ is the time constant,  $\mu_j$ is the magnitude assigned to
neuron $j$, $^j t_i^O$ is the time elapsed since the generation of the $i^{th}$
most recent efferent spike of the output neuron $j$, and $\mathcal{F}_j$ is the
set of past spikes of output neuron $j$. We observe in passing that we have
chosen the above form of the kernel for the sake of simplicity and that our
analysis applies to any differentiable kernel function with compact support.
The final force $F$ applied to the plant is
\begin{equation}
F = \sum_j \pm F_j 
\end{equation}
where $\pm$ represents the sign of $F_j$, $+$ for neurons with positive
forces and $-$ for those with negative forces (directed opposite to the
positive forces). By using groups of neurons with opposing forces, we can set
the kernels as positive functions without loss of generality.

\subsection{Gradient of the error function}

Our overall objective is to compute the gradient of the error with respect to
the synaptic weights on all the neurons in the controller network. We do this
in stages. We first compute the gradient with respect to the output spike times
of the output neurons of the controller network. Applying chain rule, we have

\begin{align} \label{eq:error}
\frac{\partial E}{\partial (^j t_l^O)}
&= \frac{\partial E}{\partial F}
\frac{\partial F}{\partial (^j t_l^O)}
\end{align}

where $^j t_l^O$ is the time elapsed since the departure of the $l^{th}$ most
recent efferent spike of output neuron $j$, and

\begin{align} \label{eq:force_grad}
\frac{\partial E}{\partial F}
=  \frac{1}{2} \sum_i \frac{\partial (x_i(0) - x_i^*(0))^2}{\partial x_i (0)} 
\frac{\partial x_i (0)}{\partial F}
=  \sum_i (x_i (0) - x_i(0)^*)
\frac{\partial x_i (0)}{\partial F}
\end{align}
and
\begin{align} \label{eq:force_grad2}
\frac{\partial F}{ \partial (^j t_l^O)}
= \frac{\partial F}{\partial F_j} 
\frac{\partial F_j}{\partial (^j t_l^O)}
= \pm \mu_j
\frac{\partial \kappa}{\partial t} \Big | _{^j t_l^O}
\end{align}

In Eq (\ref{eq:force_grad}), $\frac{\partial x_i (0)}{\partial F}$ is computed
as a numerical derivative from the plant:
$\frac{\partial x_i (0)}{\partial F} \approx \frac{\Delta x_i (0)}{\Delta F}$.

\subsection{Perturbation analysis}

Our goal now is to determine how perturbations in synaptic weights assigned to
spikes or to segments of the continuous time input signal, for any neuron in
the controller network, translate to perturbations in the times of the spikes
generated by the output neurons of the network. We achieve this by conducting a
perturbation analysis across any given individual neuron and combine it with a
recursive framework that extends the outcome of the analysis across multiple
neurons in the network. A less general analysis that applies to feedforward
networks was presented in \citep{kang2017learning}.

We begin by elucidating the notions of a direct effect as opposed to an
indirect effect of a perturbation, and in the process construct a recursive
framework for the accounting of these effects. Consider any neuron in the
network for which the weight or the time of an afferent spike or continuous
input at a synapse is perturbed. This perturbation will naturally lead to a
perturbation in the time of an output spike that is generated subsequently.
What is important to note is that this perturbation of the output spike time is
the sum total of two kinds of effects: the first is the direct effect of the
perturbation in the input spike, and the second is the indirect effect
propagated through the intermediate spikes (the spikes generated in between the
input being perturbed and the output spike under consideration) generated by
the neuron. To elaborate, a perturbation in the weight or time of an input
spike perturbs the immediately generated output spike, which in turn perturbs
the time of subsequent output spikes, and so forth. This domino effect, we
define as the indirect effect of a perturbation.  By definition, then, all
perturbations that occur via intermediate spikes, such as across two or more
neurons, are indirect effects.

Formally, we identify a direct effect with the partial derivative. So, for
example, $\frac{\partial t_l^O}{\partial w_{i,j}}$ corresponds to the how the
time of spike $t_l^O$ would change if weight $w_{i,j}$ were perturbed, if all
other spike times and weights in the network could be held fixed. In
comparison, we identify the total (sum of direct and indirect) effect with the
total derivative.  So, for example, $\frac{D t_l^O}{\partial w_{i,j}}$
corresponds to the how the time of spike $t_l^O$ would change if weight
$w_{i,j}$ were perturbed, if all perturbations in other spike times and weights
in the network due to the change in the weight $w_{i,j}$ were taken into
consideration.

The relationship between the two lends itself naturally to a recursive
formulation.

\begin{align} \label{eq:totalderiv}
\frac{D t_l^O}{\partial w_{i,j}} =
\frac{\partial t_l^O}{\partial w_{i,j}} + \sum_{k \in \Gamma_{*}} \frac{\partial t_l^O}{\partial t^O_{k}} \frac{D t^O_{k}}{\partial w_{i,j}} 
\end{align}

where $\Gamma_{*}$ is the set of all spikes generated by all neurons in the
network since the time of the spike corresponding to $w_{i,j}$. The first term
corresponds to the direct effect of the weight perturbation and the second term
corresponds to the indirect effect through other spikes. Likewise, the total
derivative of $t_l^O$ with respect to the continuous input weight
$\tilde{w}_{i,p}$ is

\begin{align} \label{eq:totalderiv-raw}
	\frac{D t_l^O}{\partial \tilde{w}_{i,p}} =
	\frac{\partial t_l^O}{\partial \tilde{w}_{i,p}} + \sum_{k \in \Gamma_{*}}
	\frac{\partial t_l^O}{\partial t^O_{k}} \frac{D t^O_{k}}{\partial \tilde{w}_{i,p}} 
\end{align}
where $\Gamma_{*}$ is the set of all spikes generated by all neurons in the
network since the segment of continuous input corresponding to
$\tilde{w}_{i,p}$.

There are only two cases where the direct effect from one spike/ segment of
continuous input to another spike is non-zero: input to output across a neuron
mediated by the change in postsynaptic potential, and output to output at a
neuron mediated by the change in the afterhyperpolarization potential. In all
other cases, the direct effect is zero. We derive the values of these next.

Consider the state of a neuron at the time of the generation of its output
spike $t_l^O$. The membrane potential of the neuron is

\begin{align} \label{eq:rnn_before_perturb} 
\tilde \Theta =
	\sum_{i \in \Gamma_c} \tilde{w}_{i,l} x_i (t_l^O)
+ \sum_{i \in \Gamma_s} \sum_{j \in \mathcal{F}_i} w_{i,j}~ \xi (t^I_{i,j} - t_l^O) 
+ \sum_{k \in \mathcal{F}} \eta ( t^O_k - t_l^O).
\end{align}

where $\Gamma_c$ is the set of synapses receiving continuous input, $\Gamma_s$
is the set of synapses receiving afferent spikes, $\tilde{w}_{i,l}$ is the
weight of synapse $i$ immediately prior to output spike $l$, and $w_{i,j}$ is
the weight of afferent spike $j$ at synapse $i$. Note that we have replaced
$\Theta$ with $\tilde \Theta$ to account for those spikes that at the time of
the generation of $t_l^O$ were less than $\Upsilon$ old, but are now past that
bound. Had the various quantities in Eq~(\ref{eq:rnn_before_perturb}) been
perturbed, we would have

\begin{align} \label{eq:rnn_after_perturb} 
\tilde \Theta &= \sum_{i \in \Gamma_c} 
	(\tilde{w}_{i,l} + \Delta \tilde{w}_{i,l}) x_i (t_l^O + \Delta t_l^O)
+ \sum_{i \in \Gamma_s} \sum_{j \in \mathcal{F}_i} (w_{i,j} + \Delta w_{i,j})
~ \xi (t^I_{i,j} + \Delta t^I_{i,j} - t_l^O - \Delta t_l^O) 				
\nonumber \\
&+ \sum_{k \in \mathcal{F}} \eta ( t^O_k + \Delta t^O_k - t_l^O - \Delta t_l^O).
\end{align}

Using a first order Taylor approximation, we get
\begin{align} \label{eq:rnn_taylor1} 
\tilde \Theta 
	&= \sum_{i \in \Gamma_c} (\tilde{w}_{i,l} + \Delta \tilde{w}_{i,l}) 
	\left( x_i (t_l^O)	+ \frac{\partial x_{i}} {\partial t}\Big |_{t_l^O}
	\Delta t_l^O \right)
\nonumber \\
&  ~ ~ ~ + \sum_{i \in \Gamma_s} \sum_{j \in \mathcal{F}_i} (w_{i,j} + \Delta w_{i,j})
	\left(\xi (t^I_{i,j} - t_l^O) + \frac {\partial \xi } {\partial t}
	\Big |_{(t^I_{i,j} - t_l^O)} ( \Delta t^I_{i,j} - \Delta t_l^O ) \right)
\nonumber \\
& ~ ~ ~ + \sum_{k \in \mathcal{F}} \left(\eta ( t^O_k - t_l^O)
	+ \frac {\partial \eta } {\partial t} \Big |_{( t^O_k - t_l^O)} 
( \Delta t^O_k - \Delta t_l^O) \right) 
\end{align} 

Combining Eq (\ref{eq:rnn_before_perturb}) and (\ref{eq:rnn_taylor1}), dropping
higher order terms and rearranging, we get

\begin{align} \label{eq:rnn-taylor2} 
        \Delta t_l^O
&= \frac{ \displaystyle
	\splitfrac{
		\sum_{i \in \Gamma_c} \Delta \tilde{w}_{i,l}  x_i (t_l^O) 
		+ \sum_{i \in \Gamma_s} \sum_{j \in \mathcal{F}_i} \Delta w_{i,j}~\xi (t^I_{i,j} - t_l^O) 
	} {
		+ \sum_{i \in \Gamma_s} \sum_{j \in \mathcal{F}_i} w_{i,j} \frac { \partial \xi} {\partial t} \Big |_{(t^I_{i,j} - t_l^O)}
		\Delta t^I_{i,j} 
		+ \sum_{k \in \mathcal{F}} \frac {\partial \eta} {\partial t} \Big |_{( t^O_k - t_l^O)}
		\Delta t^O_k 
} }
{\displaystyle
	- \sum_{i \in \Gamma_c} \tilde{w}_{i,l} 
	\frac{\partial x_{i}} {\partial t} \Big |_{t_l^O} +
	\sum_{i \in \Gamma_s} \sum_{j \in \mathcal{F}_i} w_{i,j} \frac {\partial \xi}{\partial t} \Big |_{(t^I_{i,j} - t_l^O)}
	+ \sum_{k \in \mathcal{F}} \frac {\partial \eta} {\partial t} |_{( t^O_k - t_l^O) }
}
.
\end{align}

We can now derive the final set of quantities of interest from Eq
(\ref{eq:rnn-taylor2}).  

\begin{align} \label{eq:rnn-perturb} 
	\frac {\partial t_l^O}{\partial \tilde{w}_{i,l}}
&= \frac{ \displaystyle
	x_i(t_l^O) 
}
{\displaystyle
	- \sum_{i \in \Gamma_c} w_{i,l} \frac{\partial x_{i}} {\partial t} \Big |_{t_l^O} +
	\sum_{i \in \Gamma_s} \sum_{j \in \mathcal{F}_i} w_{i,j} \frac {\partial \xi}{\partial t} \Big |_{(t^I_{i,j} - t_l^O)}
	+ \sum_{k \in \mathcal{F}} \frac {\partial \eta} {\partial t} \Big |_{( t^O_k - t_l^O) }
},
\end{align}

\begin{align} \label{eq:rnn-perturb2} 
	\frac {\partial t_l^O}{\partial \tilde{w}_{i,p}}
&= 0 \; (p > l),  
\end{align}

\begin{align} \label{eq:rnn-perturb3} 
 \frac {\partial t_l^O}{\partial t_{i,p}^I}
&= \frac{ \displaystyle
	w_{i,p} \frac { \partial \xi} {\partial t} \Big |_{(t^I_{i,p} - t_l^O)}
}
{\displaystyle
	- \sum_{i \in \Gamma_c} w_{i,l} \frac{\partial x_{i}} {\partial t} \Big |_{t_l^O} +
	\sum_{i \in \Gamma_s} \sum_{j \in \mathcal{F}_i} w_{i,j} \frac {\partial \xi}{\partial t} |_{(t^I_{i,j} - t_l^O)}
	+ \sum_{k \in \mathcal{F}} \frac {\partial \eta} {\partial t} |_{( t^O_k - t_l^O) }
},
\end{align}
 
\begin{align} \label{eq:rnn-perturb4} 
\frac {\partial t_l^O}{\partial t_{p}^O}
&= \frac{ \displaystyle
	{
	\frac {\partial \eta} {\partial t} \Big |_{( t^O_p - t_l^O)}
} }
{\displaystyle
- \sum_{i \in \Gamma_c} w_{i,l} \frac{\partial x_{i}} {\partial t} \Big |_{t_l^O} +
\sum_{i \in \Gamma_s} \sum_{j \in \mathcal{F}_i} w_{i,j} \frac {\partial \xi}{\partial t} \Big |_{(t^I_{i,j} - t_l^O)}
+ \sum_{k \in \mathcal{F}} \frac {\partial \eta} {\partial t} \Big |_{( t^O_k - t_l^O) }
},
\end{align}

and

\begin{align} \label{eq:rnn-perturb5} 
\frac {\partial t_l^O}{\partial w_{i,j}}
&= \frac{ \displaystyle
	\xi (t^I_{i,j} - t_l^O) 
}
{\displaystyle
	- \sum_{i \in \Gamma_c} w_{i,l} \frac{\partial x_{i}} {\partial t} \Big |_{t_l^O} +
	\sum_{i \in \Gamma_s} \sum_{j \in \mathcal{F}_i} w_{i,j} \frac {\partial \xi}{\partial t} \Big |_{(t^I_{i,j} - t_l^O)}
	+ \sum_{k \in \mathcal{F}} \frac {\partial \eta} {\partial t} \Big |_{( t^O_k - t_l^O) }
}.
\end{align}

\subsection{Learning rules}

Learning is accomplished via gradient descent.  As noted earlier, the synaptic
weights are updated at the times of the generation of spikes by the output
neurons of the network. The reason for this choice is that if there are no such
spikes generated, the implication is that the process variables are in a safe
range and thus no control signal is necessary. This, in turn, indicates that
there is no evidence that necessitates weight changes. Applying chain rule, we
get

\begin{align}
	\frac{\partial E}{\partial \tilde{w}_{i,p}}
&= 
\sum_{l  \in \mathcal{F}} 
\frac{\partial E}{\partial t_{l}^O} 
	\frac{D t_{l}^O}{\partial \tilde{w}_{i,p}},
\end{align}
and
\begin{align}
\frac{\partial E}{\partial w_{i,p}}
&= 
\sum_{l  \in \mathcal{F}} 
\frac{\partial E}{\partial t_{l}^O} 
\frac{D t_{l}^O}{\partial w_{i,p}}.
\end{align}

These are computed using Eq \ref{eq:error}, \ref{eq:totalderiv},
\ref{eq:totalderiv-raw}, \ref{eq:rnn-perturb}, \ref{eq:rnn-perturb2},
\ref{eq:rnn-perturb3}, \ref{eq:rnn-perturb4}, and \ref{eq:rnn-perturb5}.

Gradient descent would require

\begin{equation}
w_{i,p} \leftarrow {w_{i,p}} - \alpha \frac{\partial E}{\partial w_{i,p}} 
\end{equation} 

where $\alpha$ is the learning rate, and likewise for $\tilde{w}_{i,p}$.
Clearly, however, we can not reach into the past to
make these changes. We therefore institute a summed delayed update to the
synapse at the current time.

\begin{equation}
w_{i} \leftarrow {w_{i}} - \sum_{p \in \mathcal{F}_i }\alpha \frac{\partial E}{\partial w_{i,p}} 
\end{equation} 

and likewise for $\tilde{w}_{i}$.

\section{Experiments - Simulated Data} \label{simulated}

\begin{figure}
	\begin{subfigure}{0.49\textwidth}
		\centering
		\includegraphics[width=3in, height=2in]{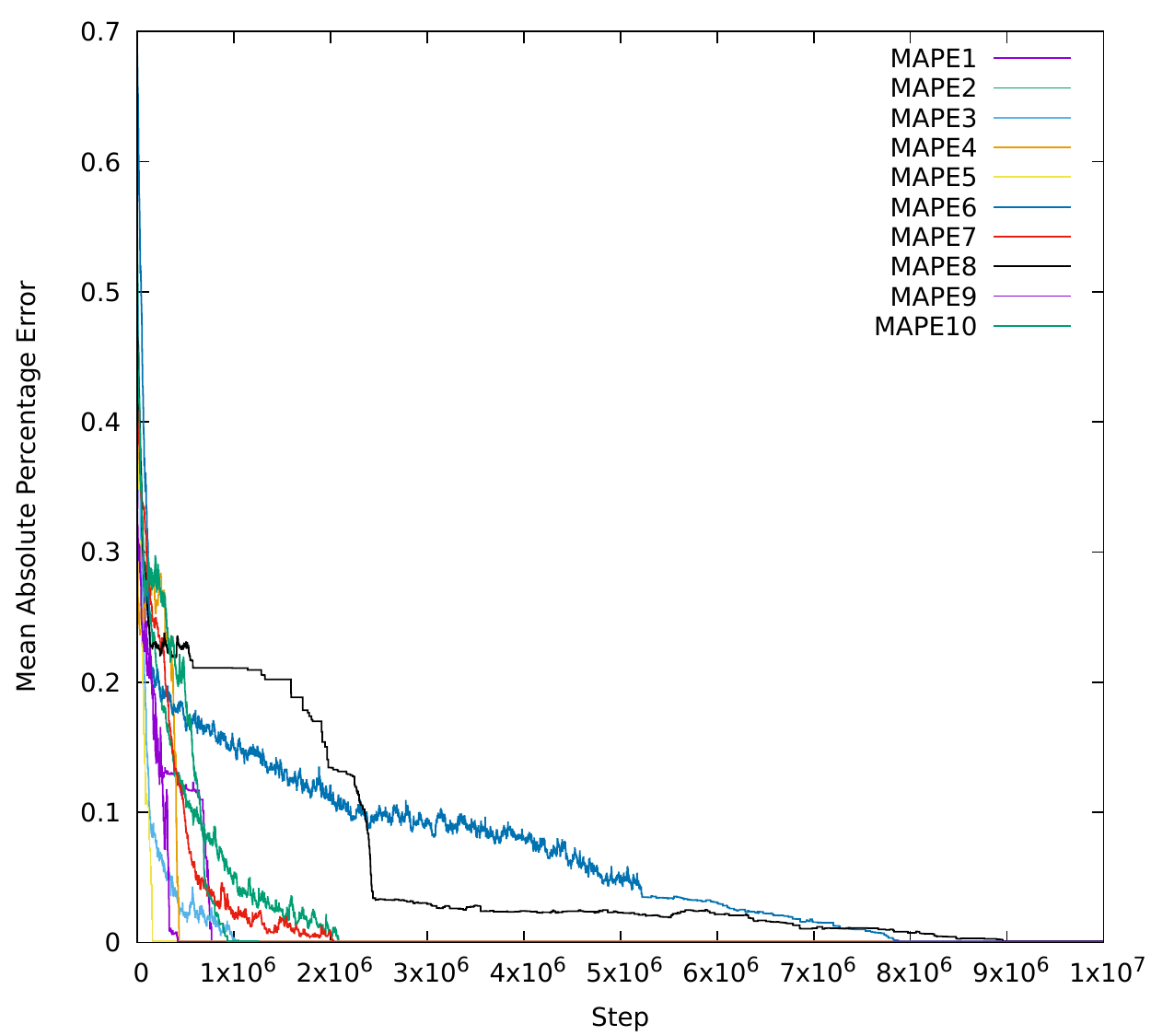}
		\caption{MAPE converges}	
		\label{convergence-mape10}
	\end{subfigure}	
	\begin{subfigure}{0.49\textwidth}
		\centering
		\includegraphics[width=3in,height=2in]{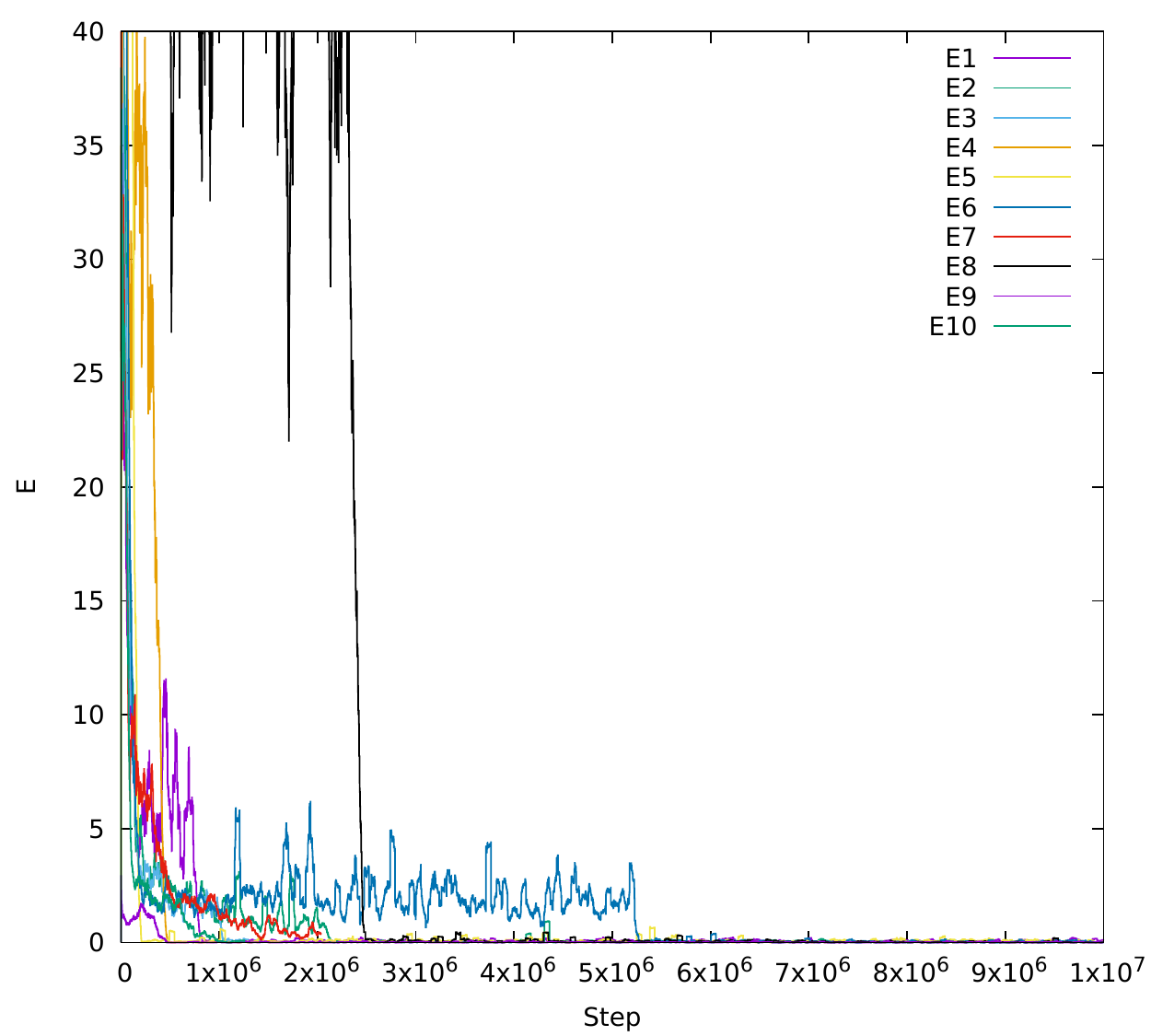}
		\caption{E converges}	
		\label{convergence-E}
	\end{subfigure}	
	\begin{subfigure}{0.49\textwidth}
		\centering
		\includegraphics[width=3in, height=2in]{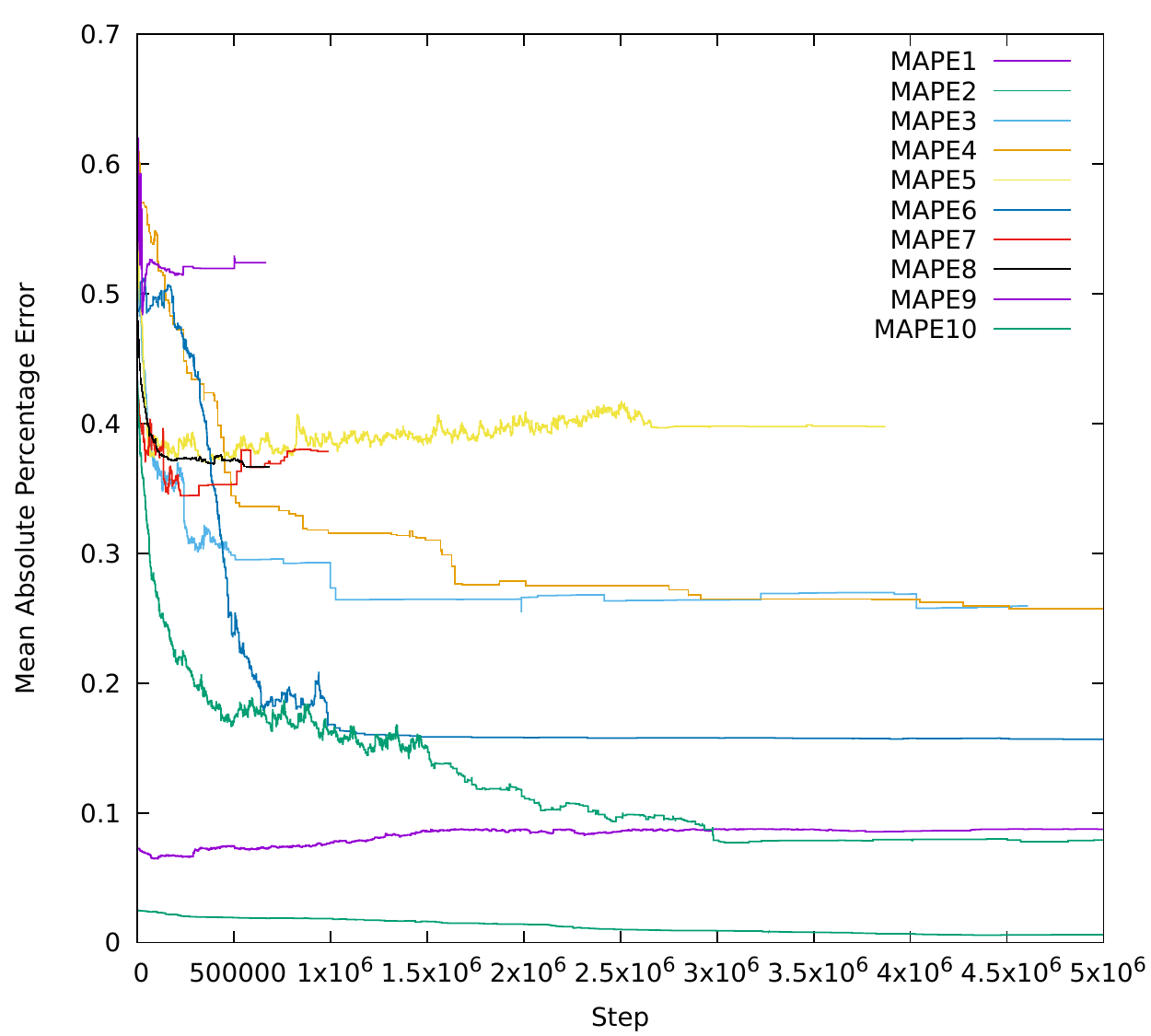}
		\caption{MAPE does not converge}	
	\end{subfigure}	
	\begin{subfigure}{0.49\textwidth}
		\centering
		\includegraphics[width=3in, height=2in]{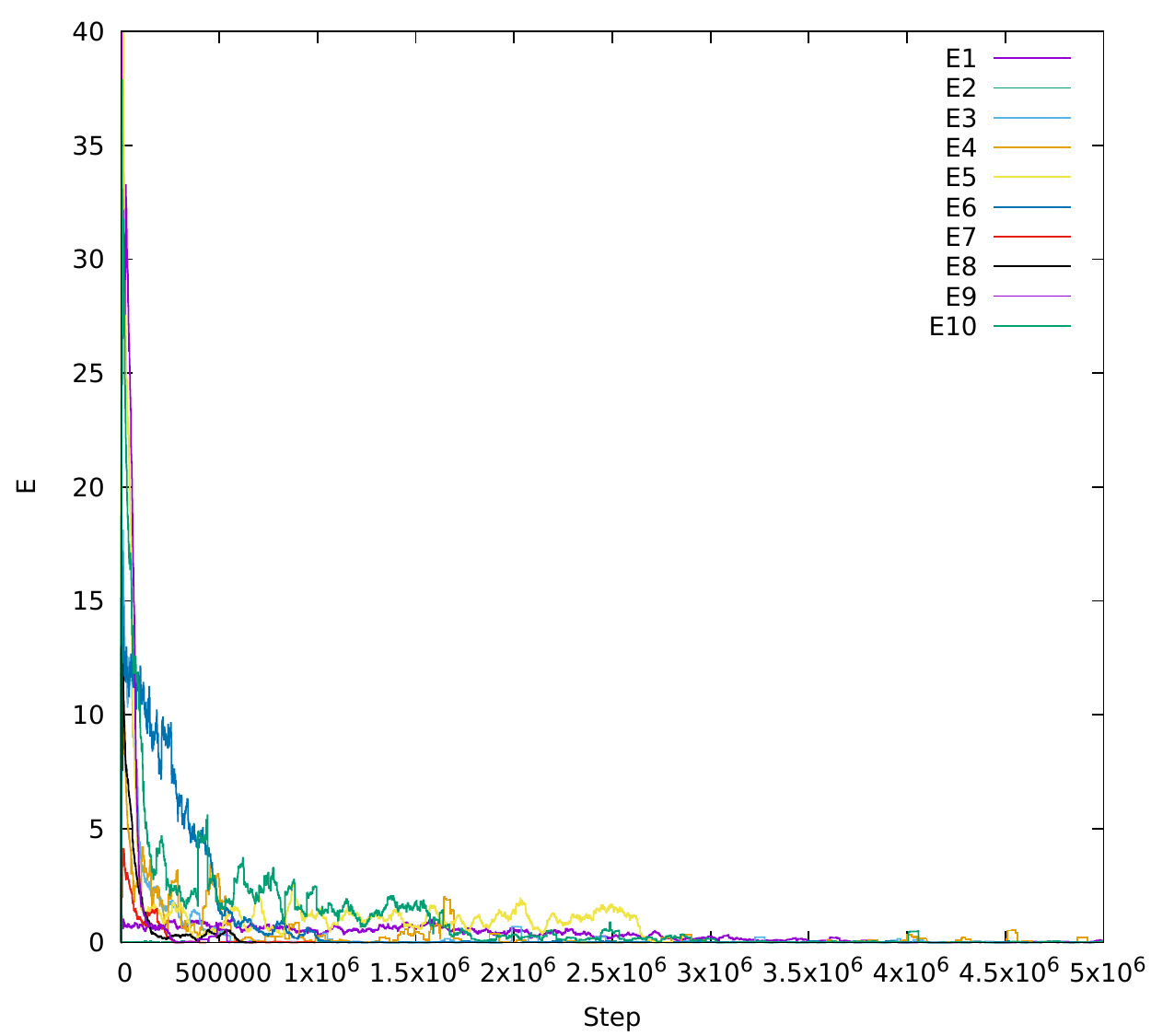}
		\caption{E converges}	
	\end{subfigure}
	\begin{subfigure}{0.49\textwidth}
		\centering
		\includegraphics[width=3in, height=2in]{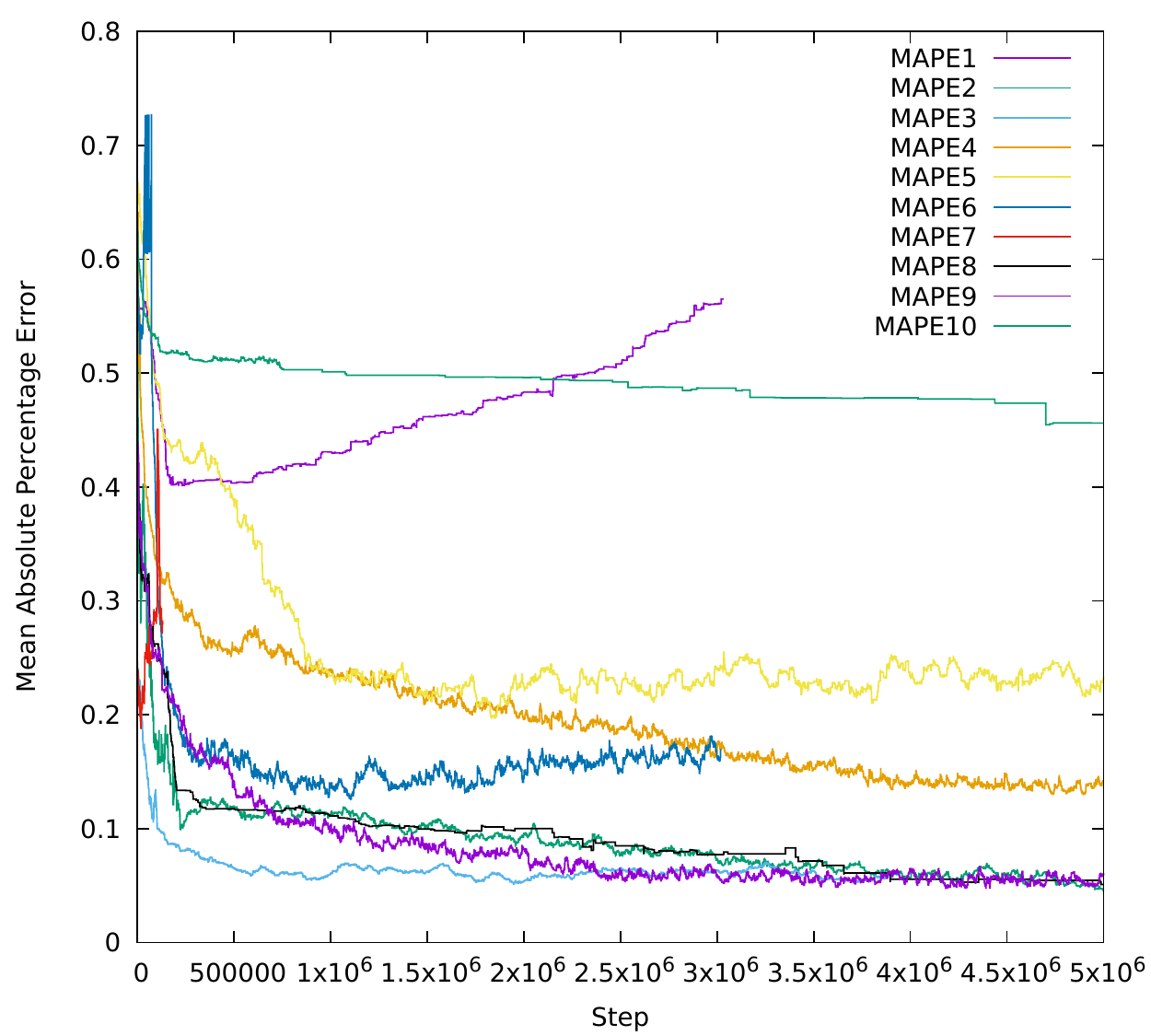}
		\caption{MAPE does not converge}	
		\label{divergedE-mape10}
	\end{subfigure}	
	\begin{subfigure}{0.49\textwidth}
		\centering
		\includegraphics[width=3in, height=2in]{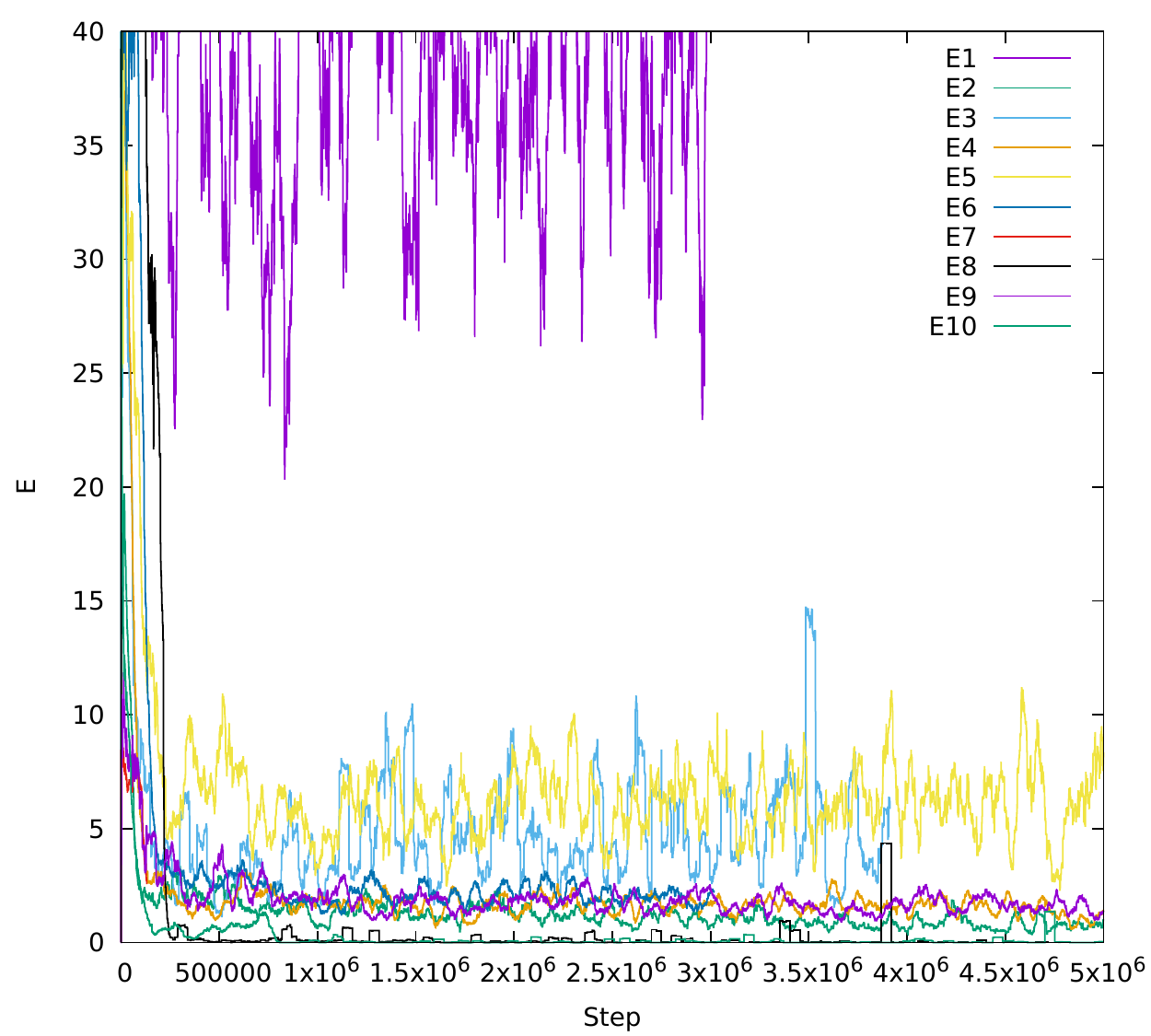}
		\caption{E does not converge}	
		\label{divergedE}
	\end{subfigure}	
	\caption{
		Convergence results for Model 3 (recurrent network) with two output
		neurons during training.	 
		The y-axis for (a), (c) and (e) corresponds to the mean absolute
		percentage error (MAPE) between the learner's weight vector and the
		putative network's weight vector.
		The y-axis for (b), (d), (f) corresponds to the error $E$.
		Captions indicate the kind of behavior observed.
	}
	\label{fig:rnn-convergence}
\end{figure}

Recognizing that recurrent networks of spiking neurons can display very complex
dynamics \cite{banerjee2006sensitive}, we begin by exploring how stochastic
gradient descent on synaptic weights influences this behavior. In particular,
we explore through simulation experiments three specific questions relevant to
the learning problem: (a) does gradient descent converge to the local minima in
spite of the complexity of the dynamics of the network, (b) is the energy
landscape non-convex, that is, does it have multiple local minimas, and (c) are
there multiple network instantiations that can effect the same control
behavior?

We addressed these questions in a highly controlled experimental setup where
the goal was for a network to learn the input output transformation of a given
\textit{putative network} of the same architecture. We considered two neuron
recurrent networks with randomly chosen synaptic weights that were driven by a
Gaussian process input signal, and whose output spike trains were convolved
against kernels that were randomly chosen and fixed. Learning networks with
randomly chosen synaptic weights were then trained using the weight update rule
presented in the previous section to learn the input output transformation of
the putative networks. Progress was measured both in terms of the error, $E$,
the squared instantaneous difference between the output of the learning versus
the putative network, and a far more conservative measure, MAPE\footnote{The
absolute value of the difference between a synaptic weight and the
corresponding synaptic weight on the putative network, normalized by the
synaptic weight on the putative network, averaged over all synapses in the
network.}, the mean absolute percentage error of the synaptic weights of the
learning network with respect to the putative network (with $100\%$
corresponding to $1.0$ in the graphs in Figure~\ref{fig:rnn-convergence}).

Figure~\ref{fig:rnn-convergence}, displays the results of a subset of the
experiments, grouped by behavior. Each panel shows the learning trajectory of
10 randomly initialized learning networks.  The $x$-axis corresponds to the
number of steps (the stepsize was set to $1 msec$), and  the $y$-axis
corresponds to either the MAPE or $E$. The convergence criterion was either
MAPE $\leq$ 0.001 or $E$ $\leq$ 0.001, depending on the scenario. The learning
rate was set to 0.0001.

Experimental results were grouped under 3 categories depending on the
convergence of MAPE and $E$: (a) and (b) where both the MAPE and $E$ converged,
(c) and (d) where the MAPE diverged but $E$ converged, and (e) and (f) where
neither the MAPE nor $E$ converged.  As can be observed from the figures, in
(a) and (b) most of the MAPEs converged within $10^7$ steps although some
networks took longer.  In the case of (c) and (d), $E$ converged but the MAPE
did not, clearly indicating that there are multiple networks that can generate
the same output trajectory when driven by identical input.  Finally, (e) and
(f) correspond to cases where neither the MAPE nor $E$ converged according to
the aforementioned criterion.  In Figure \ref{fig:rnn-convergence} (e), MAPE1,
MAPE5, and MAPE6 are clearly divergent.  The others, although convergent, did
not satisfy the criterion in the number of time steps allotted. For MAPE1,
MAPE5, and MAPE6, it is clear from the graphs that the learning rate was set
too high (the graphs display a standard thrashing behavior).

We conclude that although the $E$ landscape can have multiple local minimas for
a given set of inputs, since multiple networks can generate the same output, it
is likely that these networks can learn the necessary control objective over
time.

\section{Experiments - Inverted Pendulum} \label{pendulum}

Next, we experimented with the inverted pendulum, using the code for the plant
developed in \cite{pole}.


\subsection{Setup}

We defined a successful learning event of the controller as having balanced the
pole without failure for one hour of simulation time.  A failure was defined as
an event where the process variables of the pole departed from a certain
predefined range. Specifically, the time step for the simulations was set at $1
msec$, the number of steps for a successful run was set at 3600000 (1 hour),
and  the predefined range was [-0.2094, 0.2094] for $\theta$ and [-2.01, 2.01]
for $\dot{\theta}$, respectively.  Training of the controller was continued
until success. During the training phase, the controller was initialized with
random network weights, and the pole was initialized at a random position
$\theta$, and random angular velocity $\dot\theta$. The controller network then
learned by updating the synaptic weights as defined earlier. If a failure event
was triggered as defined above, we restarted the training with random initial
weights. Once the network had successfully learned the task, we fixed the
weights and tested the controller with random initial plant states to evaluate
its robustness.


The plant was configured as:
half-pole length $l = 0.5$ (m), pole mass $m = 0.1$ (kg), cart mass $M = 1.0$ (kg), and gravity $g = 9.8$ ($m/s^2$).
The configuration of the controller was:
time step = $1 msec$, threshold = $0.1$,
$\beta = 1.0$, $d = 1.5$, $\tau = 20$, 
$\tau_f = 20$ (msec), $R = -1000$, $\gamma = 1.2$, and $\alpha = 0.01$. 
The unit of $R$ is the same as that of the membrane potential. 
The magnitude of $R$ in the afterhyperpolarization was set large enough to
prevent inter spike intervals to fall below
$4-5 msec$.
We measured the firing rates of the output neurons in Hz (number of spikes per second).
In all the experiments, the same PID controller was used for comparison and its parameters were set as $K_p$ = 300, $K_i = 1$, and $K_d = 100$.

\subsection{Results}

We note in passing that when $\dot{\theta}$ was not provided as an input to the
network, the controller failed to learn. However, when the network received
both $\theta$ and $\dot{\theta}$, the network learned regardless of whether it
was a Model 1, 2 or 3.

\begin{figure*}
	\centering	
		\begin{subfigure}{0.45\textwidth}
			\includegraphics[width=2.6in]{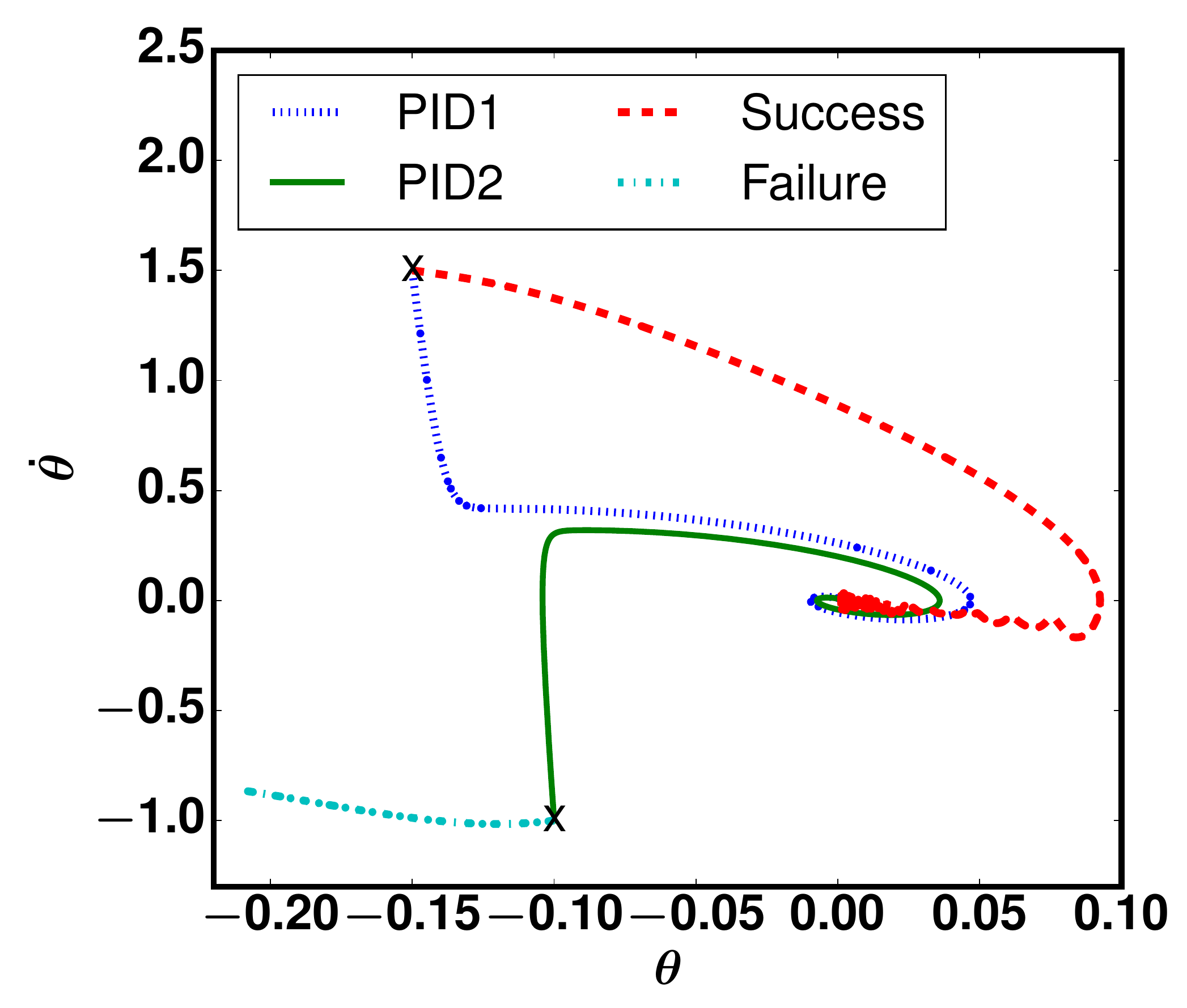}
			\caption{Trajectories}
			\label{fig:trajectories2}
		\end{subfigure}
		\begin{subfigure}{0.45\textwidth}
			\includegraphics[width=2.6in]{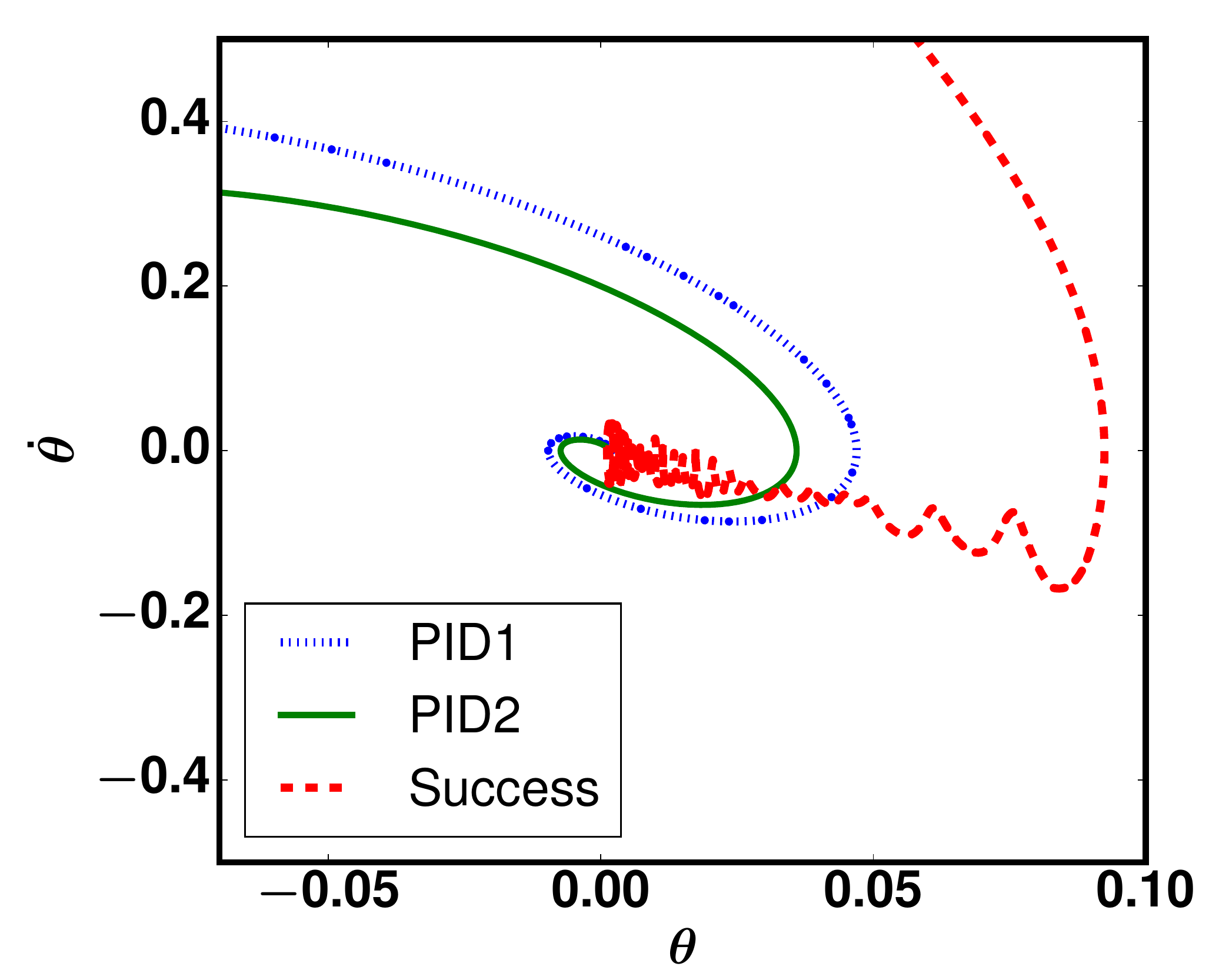}
			\caption{Zoomed in trajectories}
			\label{fig:trajectories-zoom}
		\end{subfigure}
		\begin{subfigure}{0.45\textwidth}
			\centering
			\includegraphics[width=2.0in]{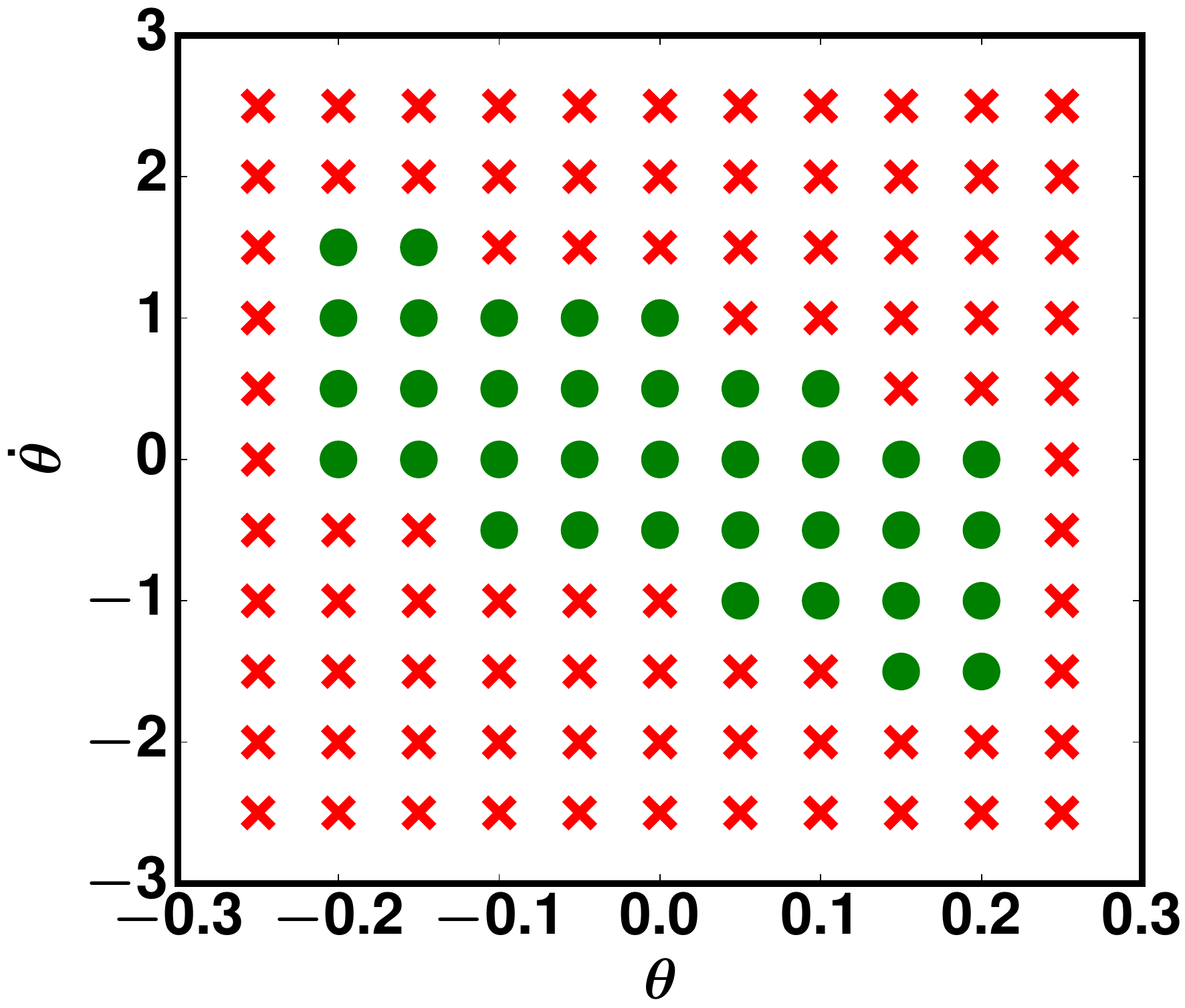}
			\caption{Coverage of Model 1}
			\label{fig:coverage2}
		\end{subfigure}
		\begin{subfigure}{0.45\textwidth}
			\centering
			\includegraphics[width=2.1in]{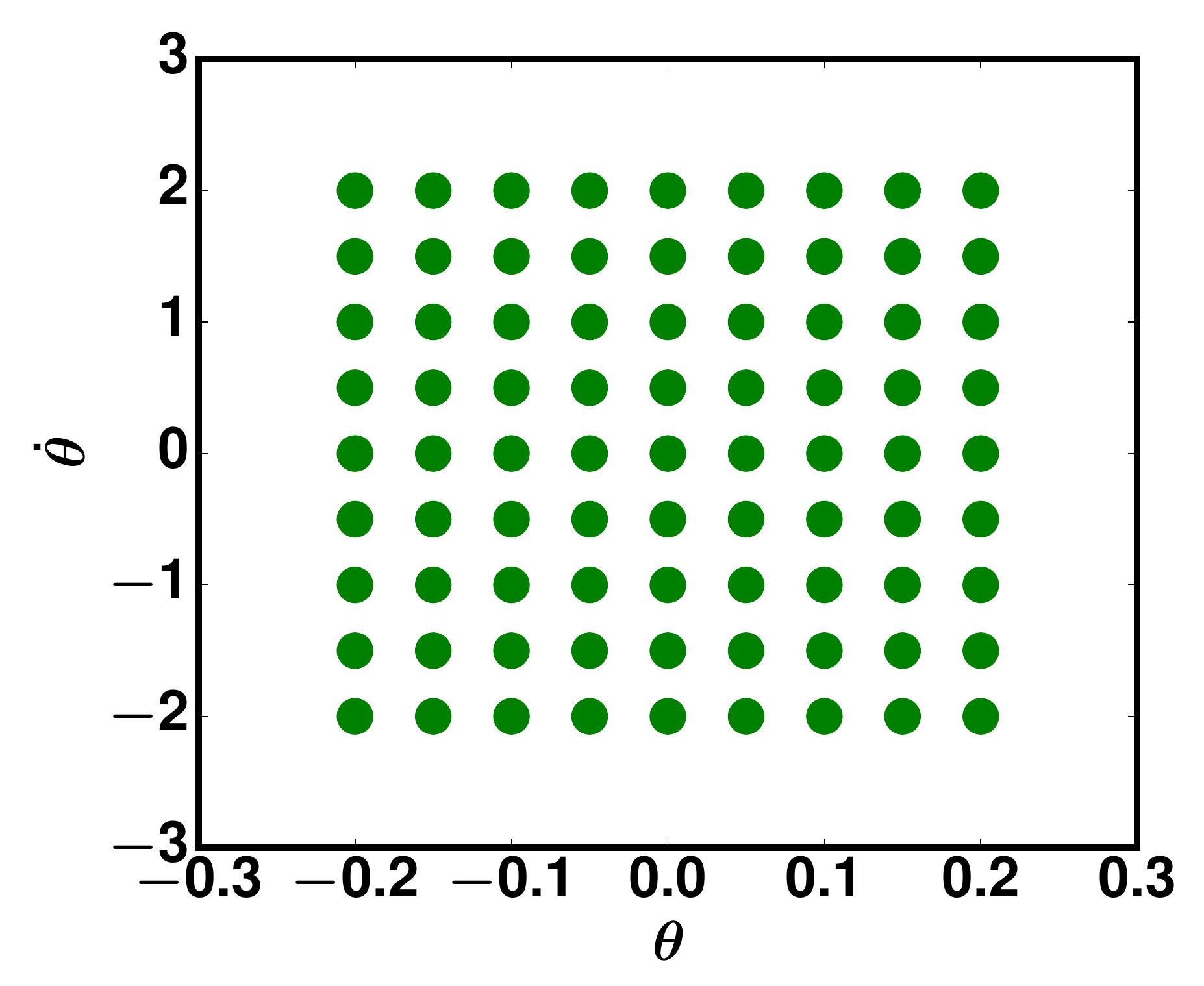}
			\caption{Coverage of PID}
			\label{fig:coverage-pid}
		\end{subfigure}
\caption{
Trajectories and coverage of Model 1.
(a) Trajectories of the plant state ($\theta$, $\dot{\theta}$) for Model 1 after training and the PID controller, over time with different initial settings.
The '$\times$' marks indicate the starting points and (0, 0) is the set point. There are two starting points, (-0.15, 1.5) where both the PID and Model 1 succeeded and (-0.1, -1.0) where the PID succeeded but Model 1 failed.
(b) Trajectories in (a) zoomed in around the set point (0, 0) for improved visualization. The plant for Model 1 oscillates between (0, -0.02) and (0, 0.02) in the stable condition.
(c) Coverage of initial states ($\theta$, $\dot{\theta}$) for the Model 1 controller. That is, initial states that the controller succeeds in shepherding to the set point (0,0). A green circle indicates a success while a red '$\times$' indicates a failure.
(d) Coverage of the PID controller. 
	}
	\label{fig:pole-train-coverage}
\end{figure*}

\subsubsection{Model 1 (Single Kernel Feedforward Network)}

The simplest controller network architecture is one where two neurons with
force kernels $+\kappa(t)$ and $-\kappa(t)$, respectively, are trained to
accomplish the control task. Surprisingly, even in this extremely restricted
scenario the network learned the task, although the domain over which the
network exercised control was found to be small relative to that of a PID
controller. Extensive details of these learning experiments have been presented
in \cite{kang2017learning}. Here we review two aspects that highlight the
nature of the resultant controller.

Fig. \ref{fig:pole-train-coverage} shows the trajectories of the plant state
($\theta$, $\dot{\theta}$) over time with two different initial settings and
the coverage over initial states of the controller, both juxtaposed against
corresponding behavior of the PID controller.  As is clear from the figure, the
trajectory of the proposed controller is qualitatively different from that of
the PID controller, demonstrating a novel control mechanism. It is also clear
that Model 1's coverage is smaller than that of the PID's.

\subsubsection{Model 2 (Multiple kernels Feedforward Network)}

The next most complex controller is one where output neurons with different
force kernels, $\pm\kappa_1(t),\ldots,\pm\kappa_N(t)$, are trained to jointly
accomplish the control task.  We performed the learning experiments above for
4, 6, and 8 output neurons with successively larger force kernels. In this
case, the force magnitude was assigned to the output neurons in a symmetric
manner: equal magnitude kernels for left and right force were assigned to pairs
of neurons.  Extensive details of these learning experiments have been
presented in \cite{kang2017learning}. Here we review the basic findings.

As in the case of Model 1, Model 2 controllers achieved the objective of
stabilizing the plant and the spike trains exhibited regular patterns; neurons
fired alternately periodically.  In general, the spike train in the stable
state was sparser than that in the start condition.  We also observed that
there were unnecessary spikes in the stable state. To elaborate, since there
are neurons that generate large as well as small forces, one only needs the
small force neurons to fire in the stable state.  This behavior can be achieved
by adding reciprocal synaptic communication between output neurons so that the
neurons are aware of each other's spike trains, which leads us to the
recurrent networks of Model 3 that we discuss in the next section.

As for the coverage of initial states that the controllers could stabilize, it
increased with the number of neurons in the controller.
When compared to the PID controller, Model 2's coverage was smaller,
albeit larger than that of Model 1. 

\subsubsection{Model 3 (Multiple kernels Recurrent Network)}

\begin{figure}
	\centering
	\begin{subfigure}{0.49\textwidth}
		\includegraphics[width=2.7in, height=2in]{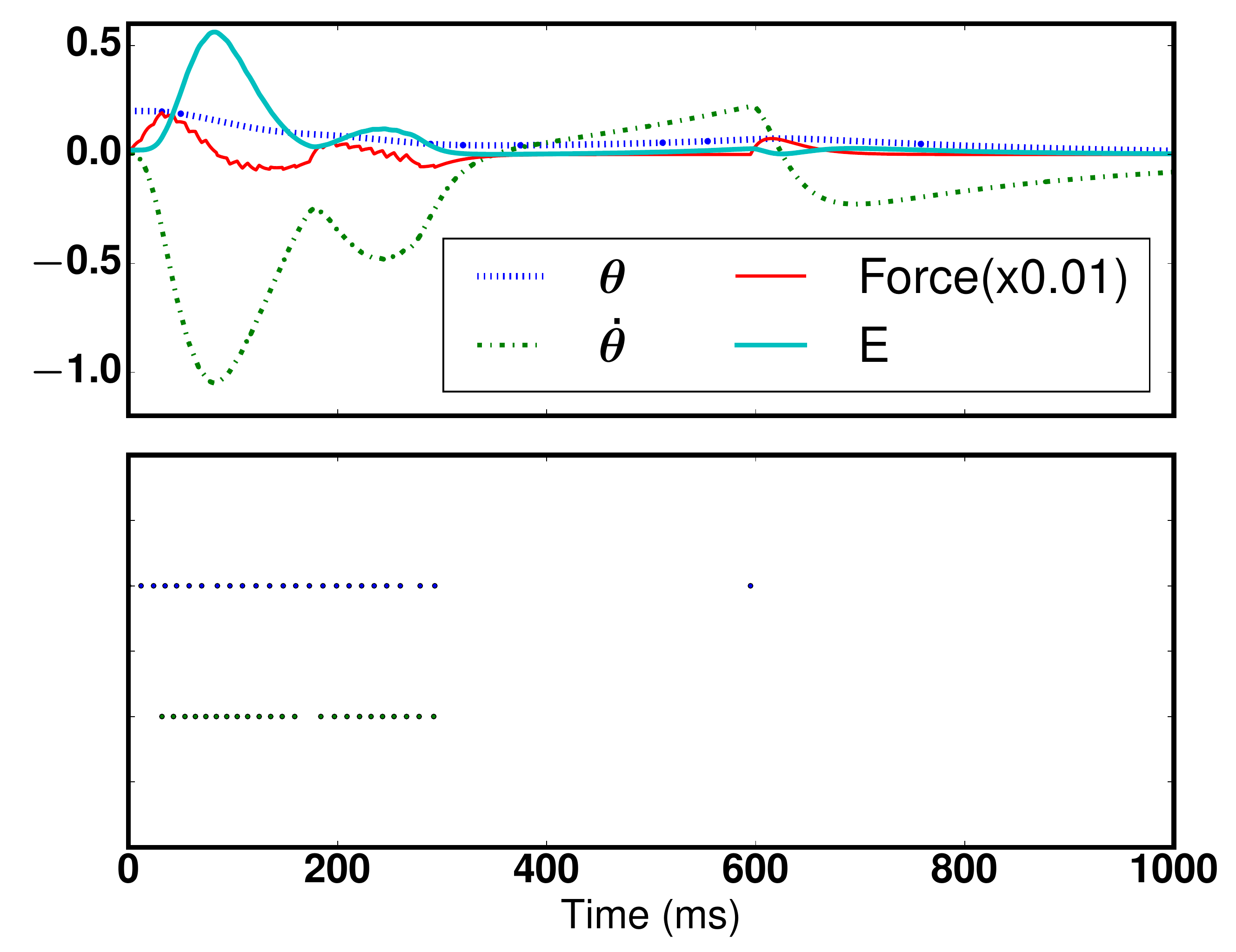}
		\caption{Starting condition}	
		\label{pole-rnn-f500-start}
	\end{subfigure}	
	\begin{subfigure}{0.49\textwidth}
		\includegraphics[width=2.7in]{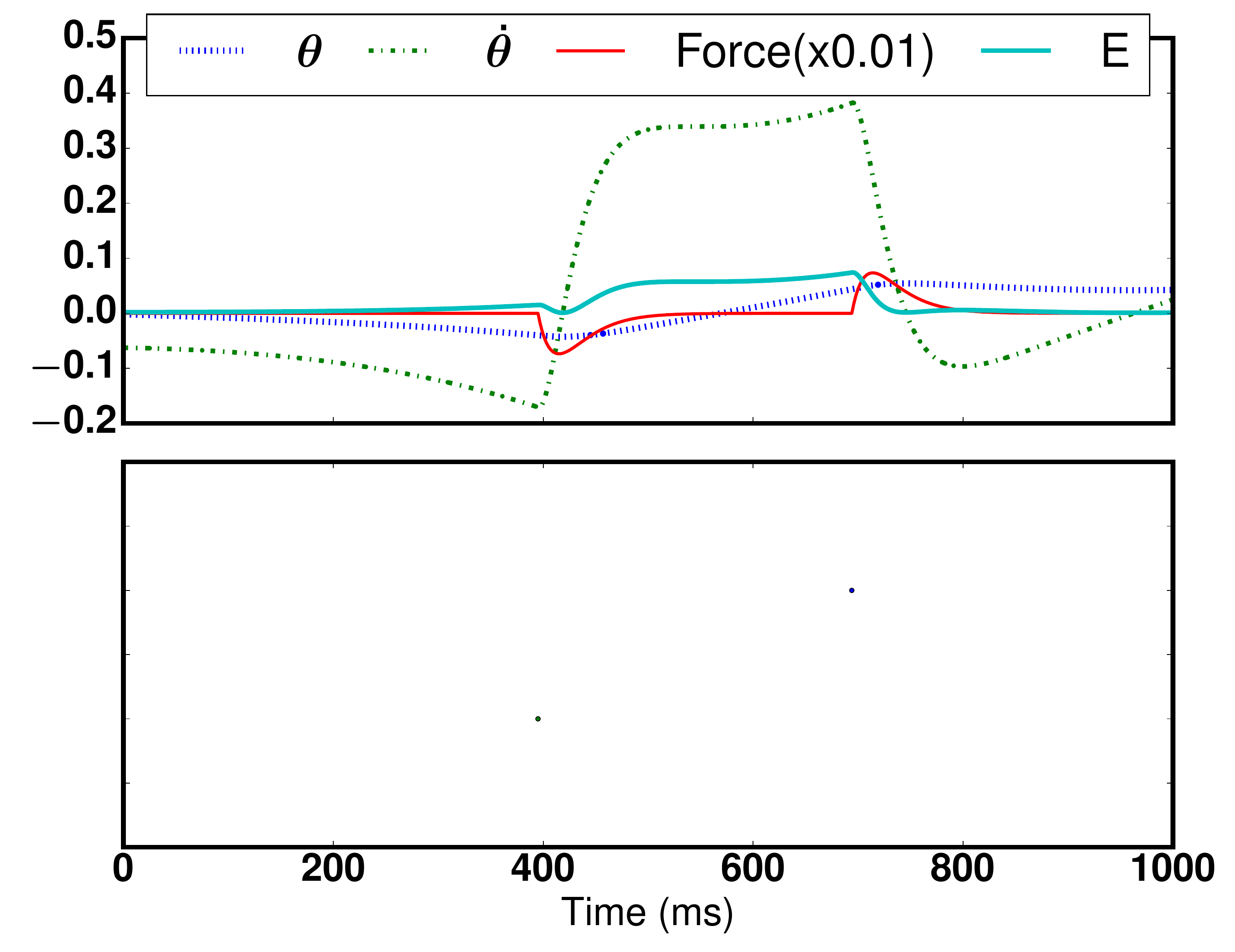}
		\caption{Stable condition}	
		\label{pole-rnn-f500-end}
	\end{subfigure}
	\caption{
		Snapshots of the plant state (top panels) and spike trains of the
		controller (bottom panels) for Model 3 with two output neurons, after training.	 
		In all the top panels, the blue dotted line is the vertical angle $\theta$, the green dash-dot is the angular velocity $\dot{\theta}$, the cyan solid line is the error function $E$, and the red solid line is the force applied to the plant.
		The final force applied to the plant is scaled down 100 fold ($\times$0.01) for improved visualization.
		(a) After training, the controller stabilizes the plant within a short
		period of time (1000 ms). The spike trains are sparse and exhibit patterns.
		(b) In the stable state, the pole oscillates around the set point (0, 0) and the error function $E$ is also at around 0.  
	}
	\label{fig:pole-rnn-snapshot}
\end{figure}

We performed the same set of learning experiments for Model 3 where the output
neurons were now recurrently connected. Fig.~\ref{fig:pole-rnn-snapshot} shows
snapshots of the plant state and the spike trains of the controller with two
output neurons, after training.  The threshold for firing was set at 0.1 and
the force magnitude was 500. Note that once the plant has been stabilized, it
remains around the set point (0, 0) and the error function $E$ also stays at
around 0.  When compared to Model 1, the spike trains were much sparser: the
average firing rates of the two output neurons were 1.03Hz and 0.52Hz,
respectively. In comparison, the firing rates for Model 1 were 33.94Hz and
34Hz.  This was the outcome of the synapses between the output neurons having
naturally converged during training to connection strengths such that when one
neuron fired, the other stopped firing, as shown in
Fig.~\ref{pole-rnn-f500-start}.

\begin{figure}
	\centering
	\begin{subfigure}{0.49\textwidth}
		\includegraphics[width=2.7in]{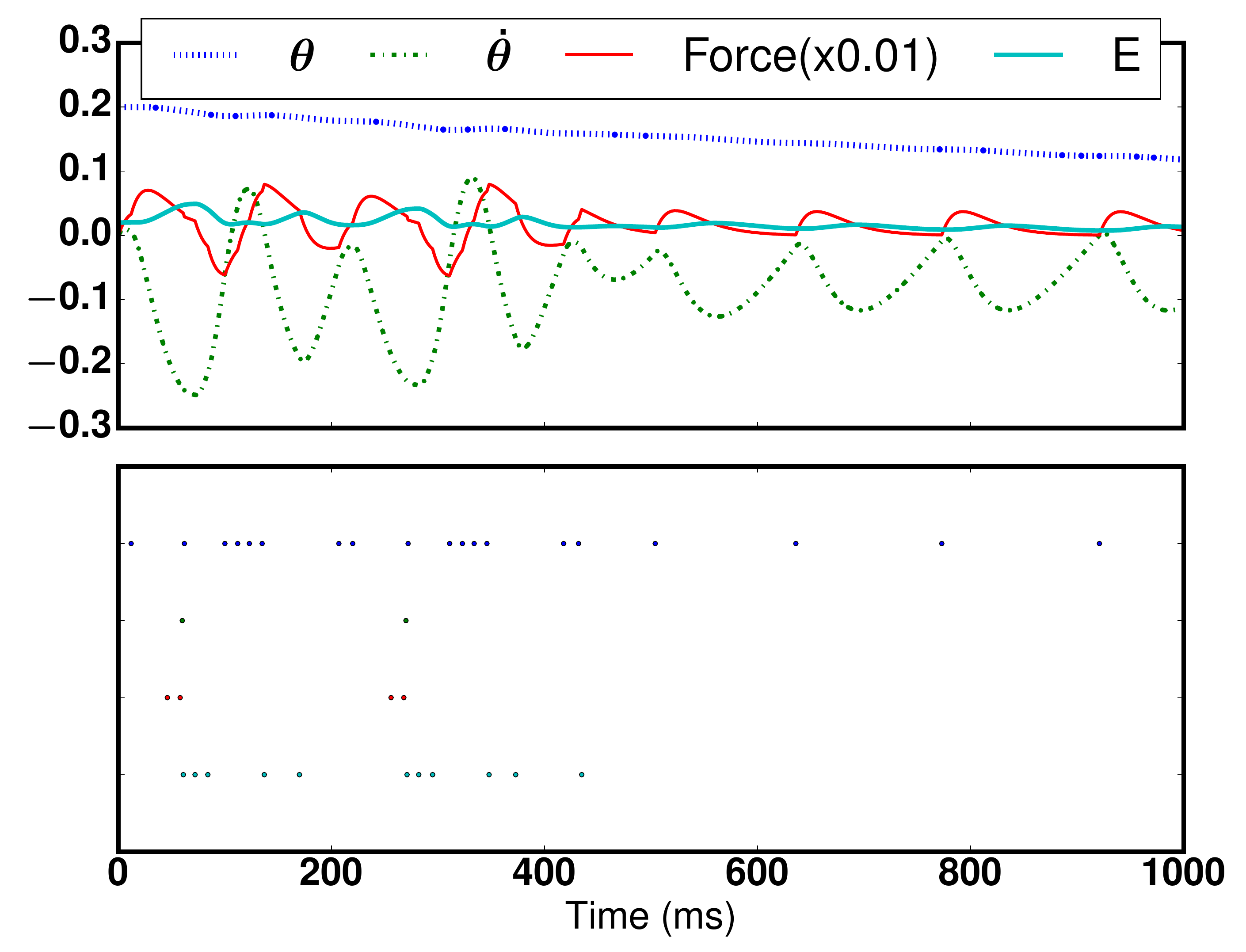}
		\caption{Starting condition for 4 output neurons}	
		\label{pole-rnn4-f300f200-start}
	\end{subfigure}	
	\begin{subfigure}{0.49\textwidth}
		\includegraphics[width=2.7in, height=2.1in]{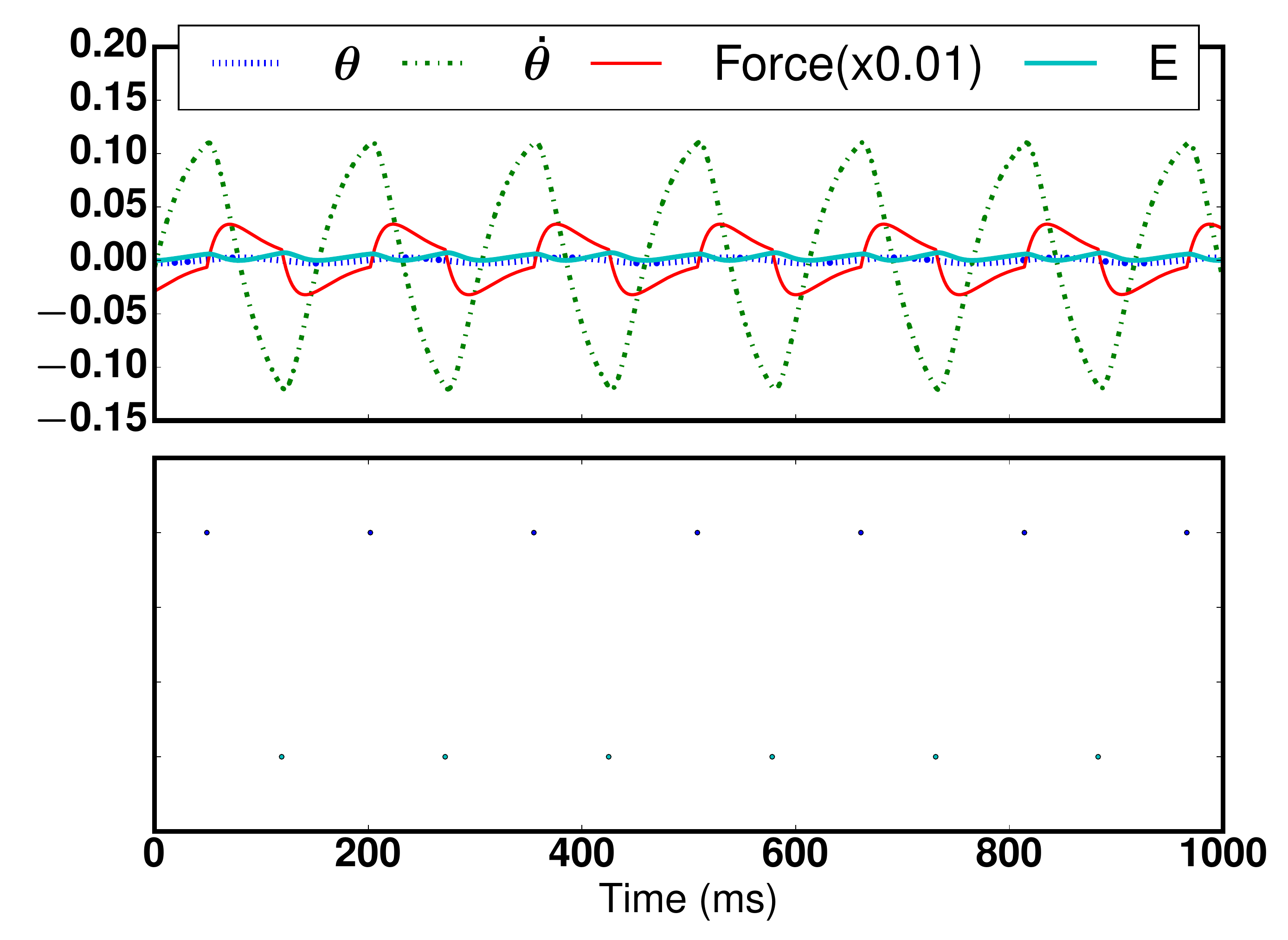}
		\caption{Stable condition for 4 output neurons}	
		\label{pole-rnn4-f300f200-end}
	\end{subfigure}
		\begin{subfigure}{0.49\textwidth}
		\includegraphics[width=2.7in]{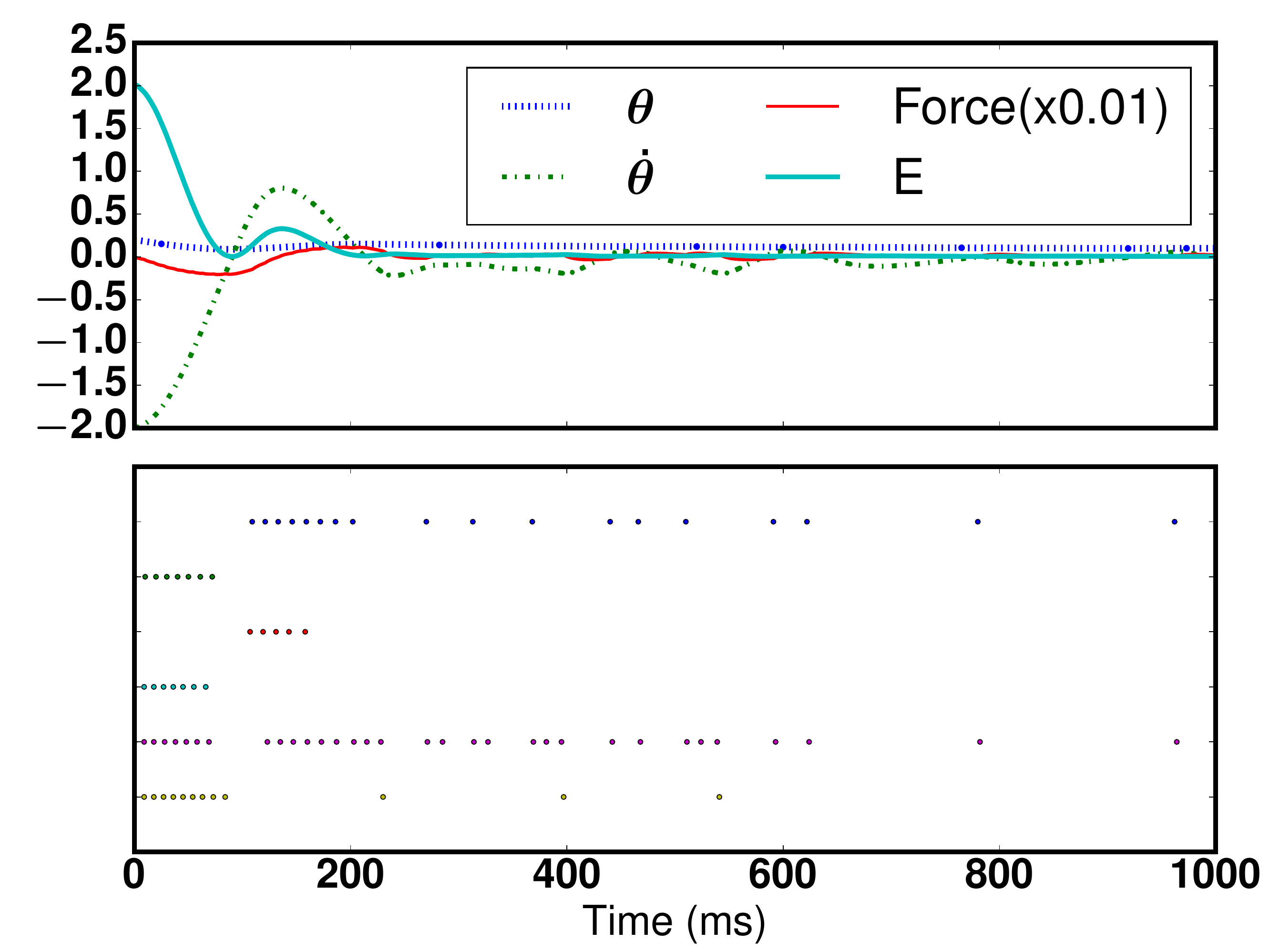}
		\caption{Starting condition for 6 output neurons}	
		\label{pole-rnn6-f300f200-start}
	\end{subfigure}	
	\begin{subfigure}{0.49\textwidth}
		\includegraphics[width=2.7in, height=2.0in]{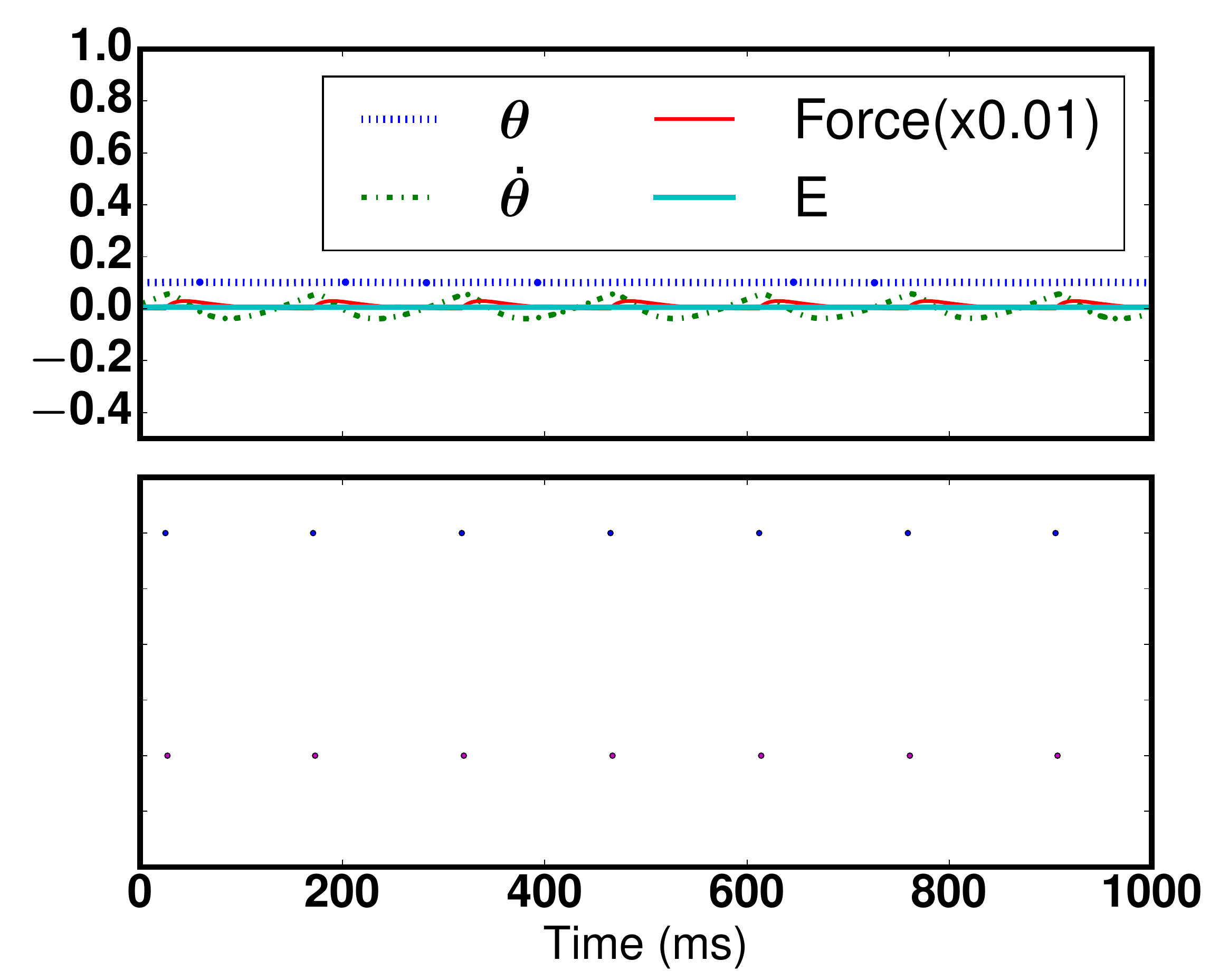}
		\caption{Stable condition for 6 output neurons}	
		\label{pole-rnn6-f300f200-end}
	\end{subfigure}
	\begin{subfigure}{0.49\textwidth}
		\includegraphics[width=2.7in]{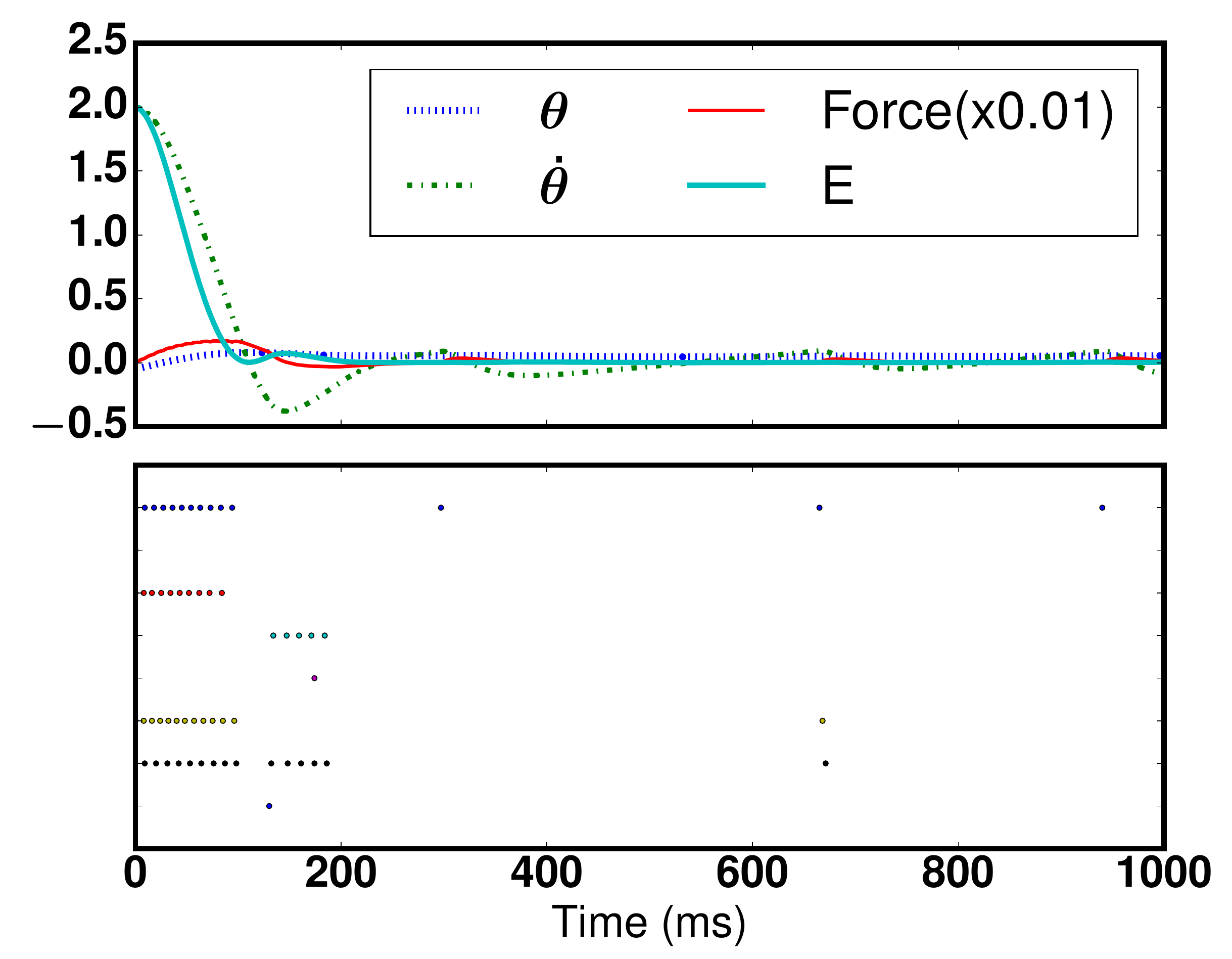}
		\caption{Starting condition for 8 output neurons}	
		\label{pole-rnn8-f300f200-start}
	\end{subfigure}	
	\begin{subfigure}{0.49\textwidth}
		\includegraphics[width=2.7in, height=2.1in]{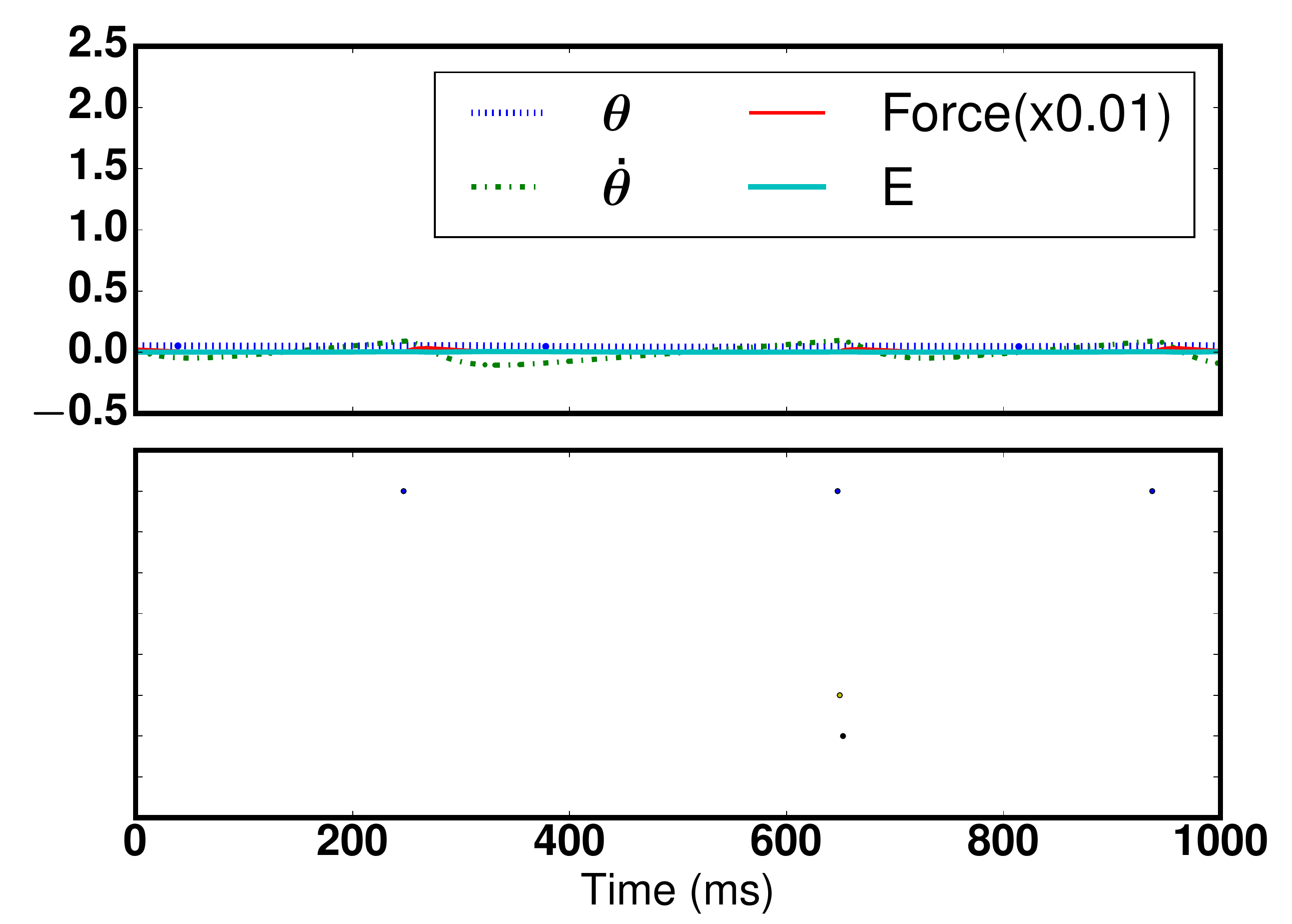}
		\caption{Stable condition for 8 output neurons}	
		\label{pole-rnn8-f300f200-end}
	\end{subfigure}
	\caption{
		Snapshots of the plant state (top panels) and the spike trains of the
		controller (bottom panels) for Model 3 with recurrently connected 4, 6,
		and 8 output neurons, after training.
		The spike trains in all cases are sparse (0.9-3.3Hz per neuron).
	}
	\label{fig:pole-rnn-multiple-snapshot}
\end{figure}

\begin{figure}
	\centering
	\begin{subfigure}{0.49\textwidth}
		\centering
		\includegraphics[width=2.0in]{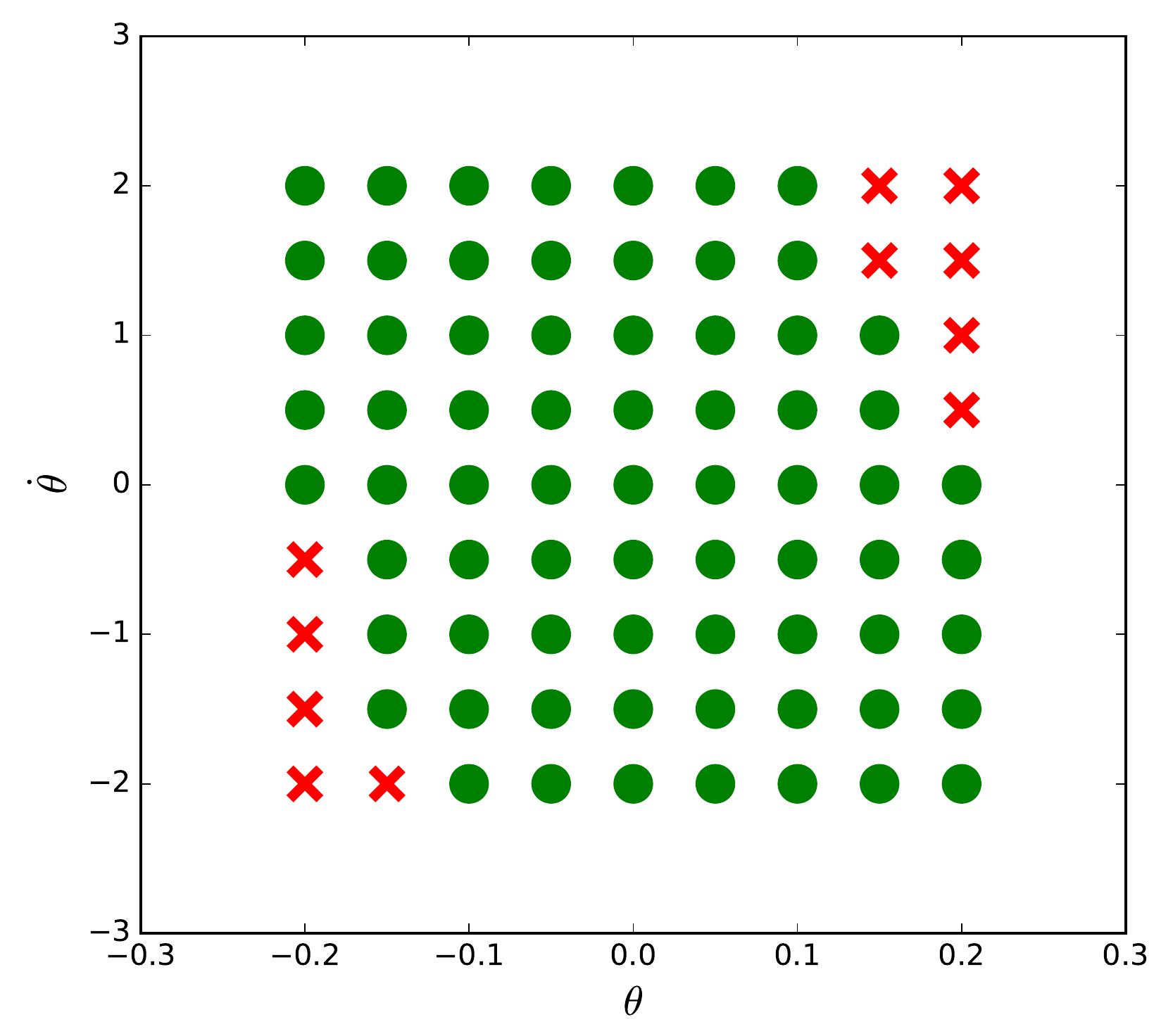}
		\caption{2 neurons}	
		\label{pole-rnn2-coverage}
	\end{subfigure}	
	\begin{subfigure}{0.49\textwidth}
		\centering
		\includegraphics[width=2.0in]{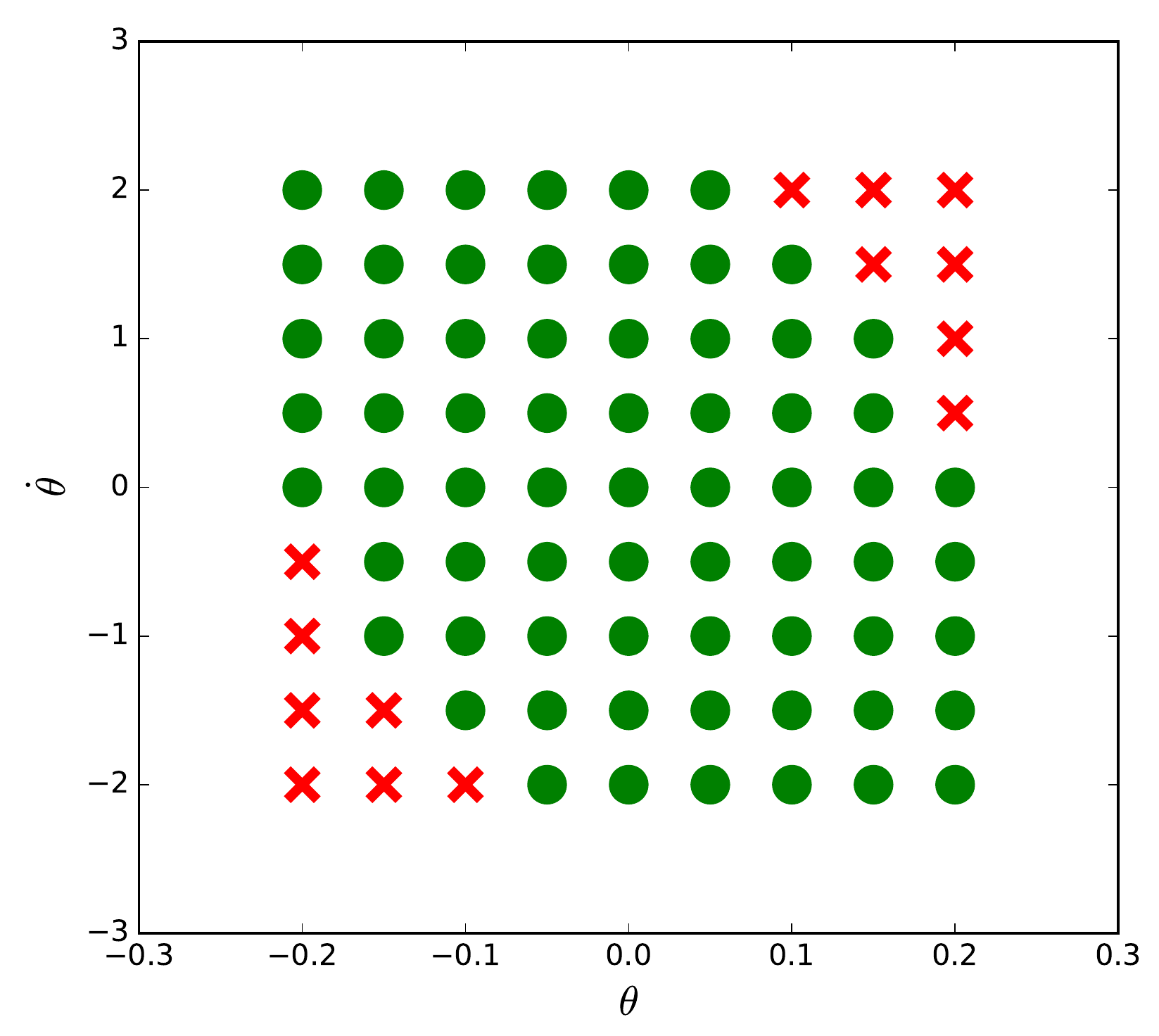}
		\caption{4 neurons}	
		\label{pole-rnn4-coverage}
	\end{subfigure}
	\begin{subfigure}{0.49\textwidth}
		\centering
		\includegraphics[width=2.0in]{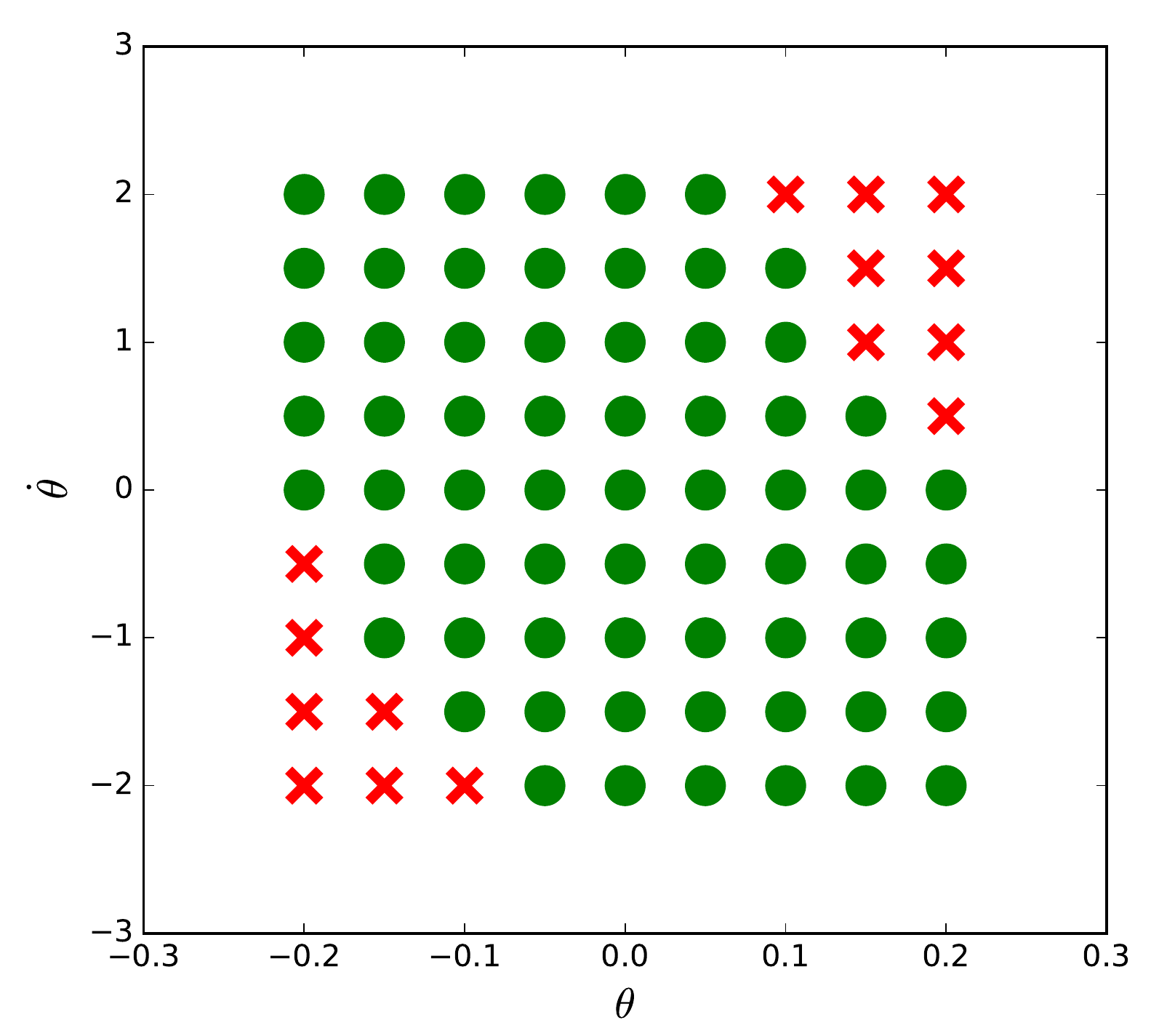}
		\caption{6 neurons}	
		\label{pole-rnn6-coverage}
	\end{subfigure}	
	\begin{subfigure}{0.49\textwidth}
		\centering
		\includegraphics[width=2.0in]{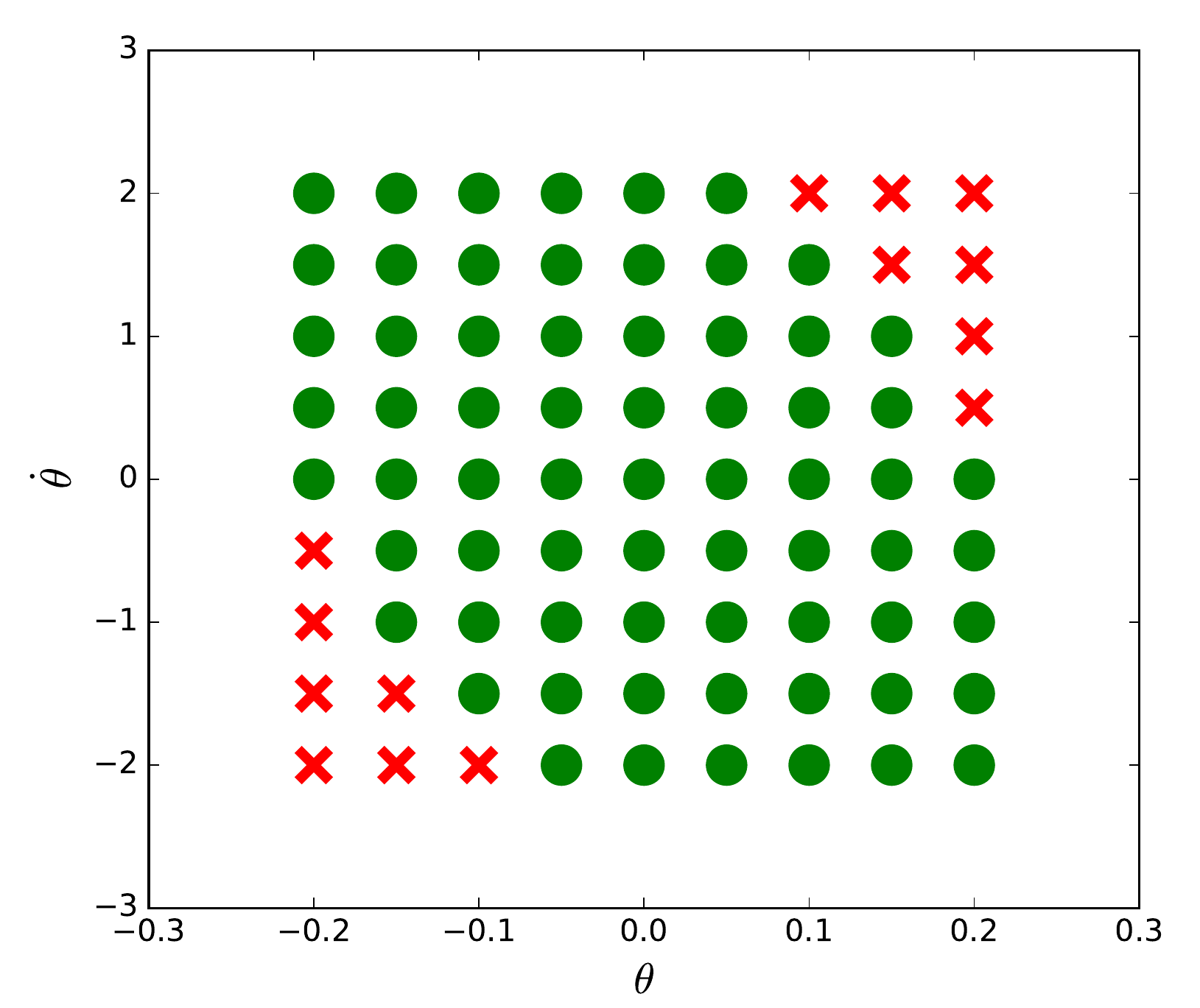}
		\caption{8 neurons}	
		\label{pole-rnn8-coverage}
	\end{subfigure}
	\caption{
	 	Stability coverages of initial plant states ($\theta$, $\dot{\theta}$)
		for Model 3 with 2, 4, 6, and 8 output neurons, after training.	 
	 	A green circle indicates success while a red '$\times$' indicates
		failure for the corresponding initial state.
		(a) 2 output neurons 
		covers 70 initial plant states among 81 attempted states (86.42\%).
		This is larger than the 36 states (44.44\%) achieved by Model 1.
		(b) 4 output neurons 
		covers 67 states (82.72\%). This is larger than the 45 states (55.56\%)
		achieved by Model 2. 
		(c) 6 output neurons 
		covers 66 states (81.48\%) compared to 51 (62.96\%) of Model 2.
		(d) 8 output neurons 
		covers 67 states (82.72\%) compared to 62 (76.54\%) of Model 2.
		Model 3 results are comparable to the PID controller which covers all
		81 states (100\%).
	}
	\label{fig:pole-rnn-coverage}
\end{figure}

We also conducted experiments with 4, 6, and 8 recurrently connected output
neuron controllers. Fig.  \ref{fig:pole-rnn-multiple-snapshot} displays
snapshots of their starting and stable conditions.  The force magnitudes
assigned to the 4 output neurons were -500, -10, 10, and 500.  The average
firing rate of the 4 output neurons combined was 13.18 Hz, shared between 6.6
Hz for the left neurons and 6.58 Hz for the right neurons. The firing rate per
neuron was 3.3Hz.  Fig.  \ref{pole-rnn4-f300f200-start} shows that after
training, the controllers stabilized the plant within a short period of time
(1000 ms).  Likewise, the average firing rate of the 6 output neurons
controller was 13.05Hz (6.53, 6.52 Hz, for left and right neurons
respectively), and the rate per neuron was 2.18Hz.  For the 8 neuron
controller, the average firing rate as shown in Fig.
\ref{fig:pole-rnn-multiple-snapshot} (e) and (f) was 7.21Hz (3.14Hz left,
4.07Hz right) and the rate per neuron was 0.9Hz. Finally,
Fig.~\ref{fig:pole-rnn-coverage} shows the coverage of initial states
($\theta$, $\dot{\theta}$) that the Model 3 controllers with 2, 4, 6, and 8
output neurons managed to stabilize.


Interestingly, the reason why the spiking network controllers could not learn
the corner cases was that there was very little time before the pendulum fell.
The PID controller, not beholden to a learning process, does not suffer this
problem. The very limited set of scenarios where the network controllers failed
to learn is testament to the versatility of the learning process.



\section{Experiments - Fish Locomotion} \label{fish}

To further explore the capacity of the proposed controller, we considered the
challenging task of locomotion. We used a fish plant with detailed emulation of
musculature (the SOLEIL project \citep{soleil}).  The difficulty of this
problem is apparent from the observation that unless the multiple controls,
$yaw/pitch/roll$ are operated synergistically, the task is impossible to
achieve. Furthermore, vastly different control trajectories may be necessary
for what would otherwise look like similar states, such as when the fish at the
same location is oriented away from, instead of toward, the target. We used a
simple routine mechanism to control the velocity of the fish and focused on the
task of learning the $yaw/pitch/roll$ controls. To elaborate, the velocity of
the fish was set to a certain value, stay constant, and then ramp down to $0$
when the target location fell within a threshold distance from the fish's
center of gravity ({\em cog}).

\subsection{Setup}

In all experiments, a fish was positioned in 3-dimensional space with initial
location of its cog set as the origin (0, 0, 0), and its head-to-tail axis
oriented along the negative y-axis (facing negative y, with tail toward
positive y).  The fish moved towards a target location $(x^*, y^*, z^*)$
according to the velocity profile described in the previous section. The fish
therefore stopped only when the controller had achieved its objective. The
locomotion of the fish was controlled by a vector of $\langle yaw, pitch,
roll\rangle$ control forces as depicted in Figure~\ref{fig:plant}b. The process
variable input into the controller was the angle between the front to back axis
of the fish and the fish centric vector pointing to the target location.  We
denote this angle vector process variable between the fish's current
orientation and the target by $\langle \theta_x, \theta_y, \theta_z \rangle $.
In the learning experiments, the controller took as input the difference
between the current angle vector $\langle \theta_x, \theta_y, \theta_z\rangle $
and the desired angle vector $\langle \theta_x^*, \theta^*_y,
\theta_z^*\rangle$, which was set at $\langle 0, 0, 0\rangle$, and output a
vector of $\langle yaw, pitch, roll\rangle$ forces.  The desired angle vector
$\langle 0, 0, 0\rangle$ corresponded to the fish facing the target head on.
The error function $E$ was therefore $\frac{1}{2}\sum_{i \in {\lbrace x,y,z
\rbrace }} (\theta_i^*-\theta_i)^2$.  For the simulations, we defined a
successful learning event as one where the fish reached the target location.
Formally, a success is defined as an event where the Euclidean distance $D$
between the fish and the target is within a certain predefined threshold.  We
conducted two sets of experiments, one where the fish was constrained on a
2-dimensional plane containing the target (all locations here being $(x,y,0)$)
and one where the fish moved in the full 3-dimensional space. We set the
threshold $D$ at 0.06 for the 2-dimensional experiments and 0.1 for the
3-dimensional experiments.  With this setup, training of the controller was
continued until a success. In the training phase, the controller started with
random network weights and a randomly chosen target location; the fish started
at the origin $(0, 0, 0)$.  The controller then learned to control the
locomotion of the fish by updating the synaptic weights.  Unless stated
otherwise, the training method and configurations of the controllers were the
same as those of the inverted pendulum.  We note in passing that there is an
upper limit to the force magnitude that can be applied to the plant. When the
force magnitude crosses this bound, the fish moves unrealistically.  Unless
otherwise stated, the force magnitude assigned to any output neuron was 50.

\subsection{Results}
We present experimental results in the 2-dimensional case for two controller
network architectures, feedforward followed by recurrent.

\subsubsection{Model 1 (Feedforward Network)}

\begin{figure}
	\centering
	\includegraphics[width=4.5in]{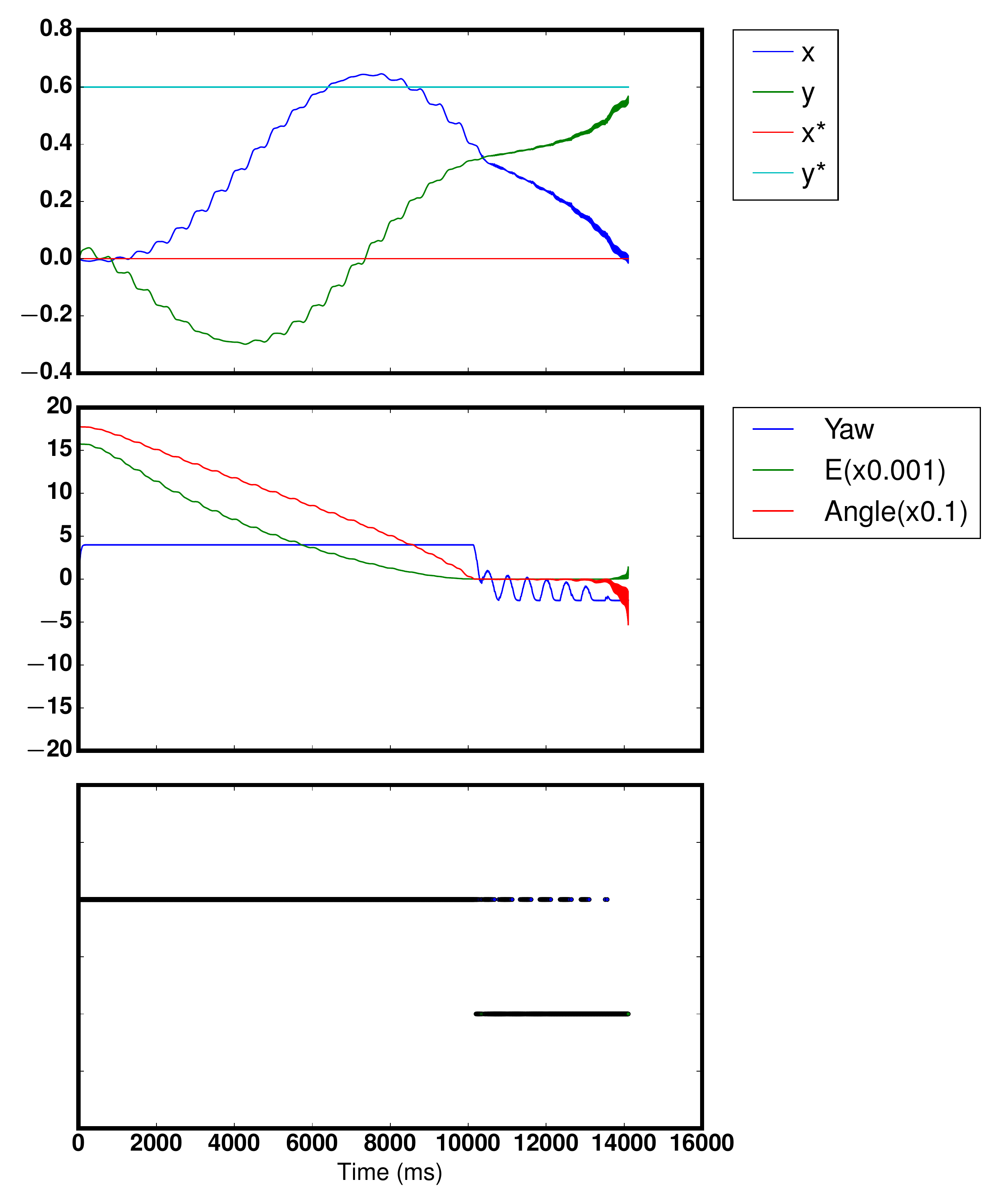}
	\caption{
		Snapshot of the system state and spike trains of the controller for Model 1 with 2 output neurons after training. The control variable $\langle yaw \rangle$ was controlled. The target location $(x^*, y^*, z^*)$ is (0, 0.6, 0). 
		In the top panel, the blue dotted line is the $x$-coordinate of the
		fish's cog, the green dash-dot is the $y$-coordinate, and the red solid
		line is the $z$-coordinate. In the middle panel, shown are the error function $E$, the distance $D$ between the fish and the target, and the $\langle yaw \rangle$ force applied to the fish, respectively.
		In the bottom panel, the spike trains of the output neurons corresponding to $\langle +yaw, -yaw \rangle$ forces are displayed.
	}
	\label{fig:fish-fnn2}
\end{figure}

We first used a feedforward network of neurons to learn the control task when
the fish was restricted to a 2-dimensional plane containing the target
location.  Note that the task was, in theory, achievable with the control of
just the {\em yaw}, and therefore we considered a Model 1 network with two
neurons forcing yaw left or right. The network learned the task with ease.
Fig.~\ref{fig:fish-fnn2} shows snapshots of the system state and spike trains
of the controller for Model 1 with 2 output neurons, after training.


Observe that when the fish gets close to the destination in terms of angular
alignment, the neurons stop generating spikes.  The dense spike trains in the
bottom panel are due to the practical limits on the force magnitude that can be
applied to the plant at one time, as mentioned earlier. If this force is too
large, the fish spins around unrealistically.  Since the network is not allowed
to generate a large force at once, in contrast to the case of the inverted
pendulum controller, the neurons generate multiple spikes to compensate.

\subsubsection{Model 3 (Recurrent Network)}

\begin{figure}
	\centering
			\includegraphics[width=4.5in]{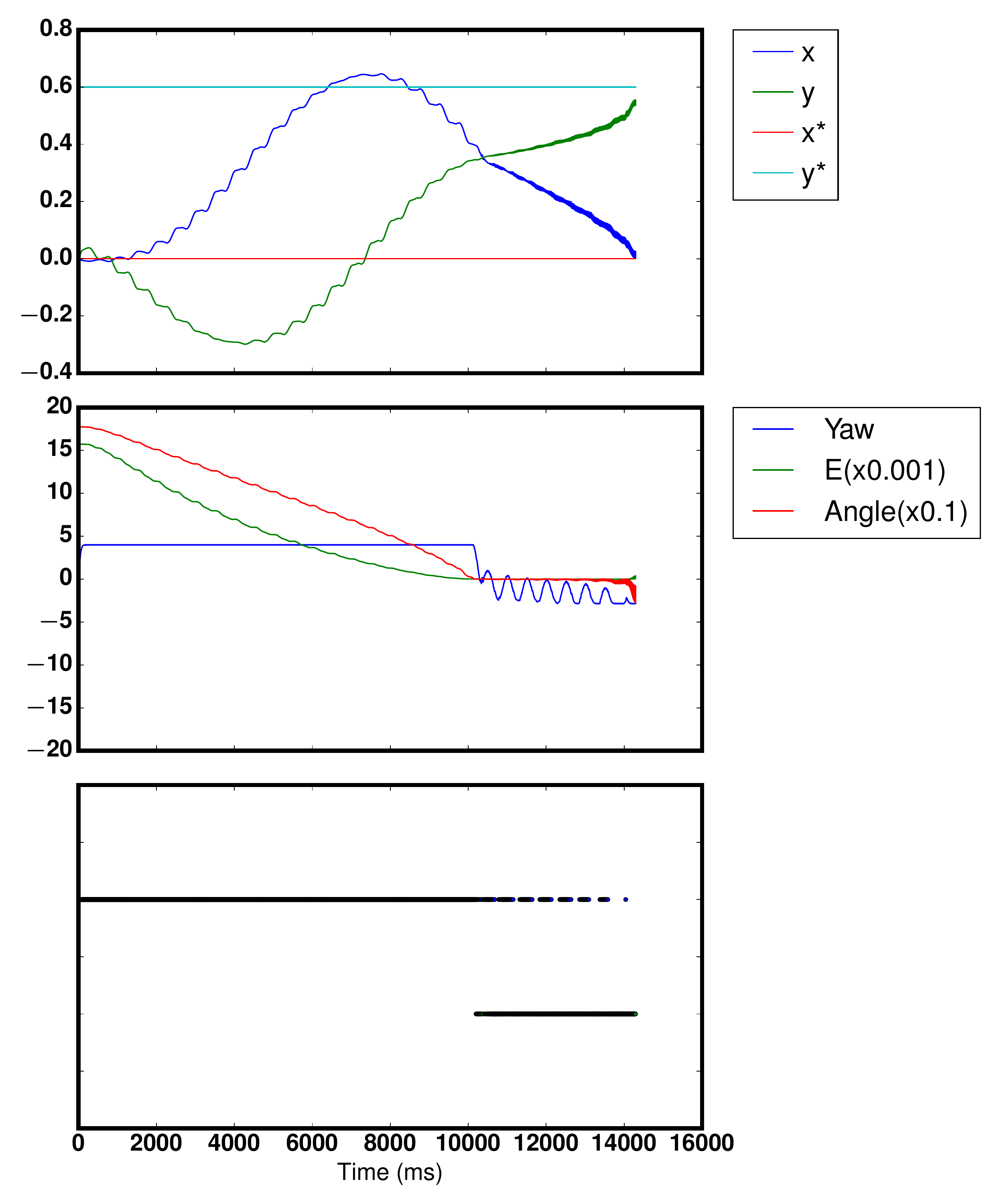}
	\caption{
		Snapshot of the system state and spike trains of the controller for
		Model 3 with 2 output neurons after training. The control variable
		$yaw$ was controlled. Note that the results are almost identical to
		that of the feedforward network with 2 neurons.
		The target location $(x^*, y^*, z^*)$ is (0, 0.6, 0). 
		In the top panel, the blue line is the $x$-coordinate of the fish's cog,
		the green line is the $y$-coordinate, and the red line is the
		$z$-coordinate. In the middle panel, shown are the error function $E$,
		the distance $D$ between the fish and the target, and the
		$\langle yaw \rangle$ force applied to the fish, respectively.
		In the bottom panel, the spike trains of the output neurons
		corresponding to $\langle +yaw, -yaw \rangle$ forces are displayed.
	}
	\label{fig:fish-rnn2}
\end{figure}

We repeated the 2-dimensional experiments with a recurrent network of two
output neurons trained to achive the same control task. Training was successful
here as well.  Fig.~\ref{fig:fish-rnn2} shows snapshot of the system state and
spike trains of the controller for Model 3 with 2 output neurons, after
training.

\begin{figure}
	\centering
	\begin{subfigure}{0.49\textwidth}
		\includegraphics[width=2.5in,height=2.2in]{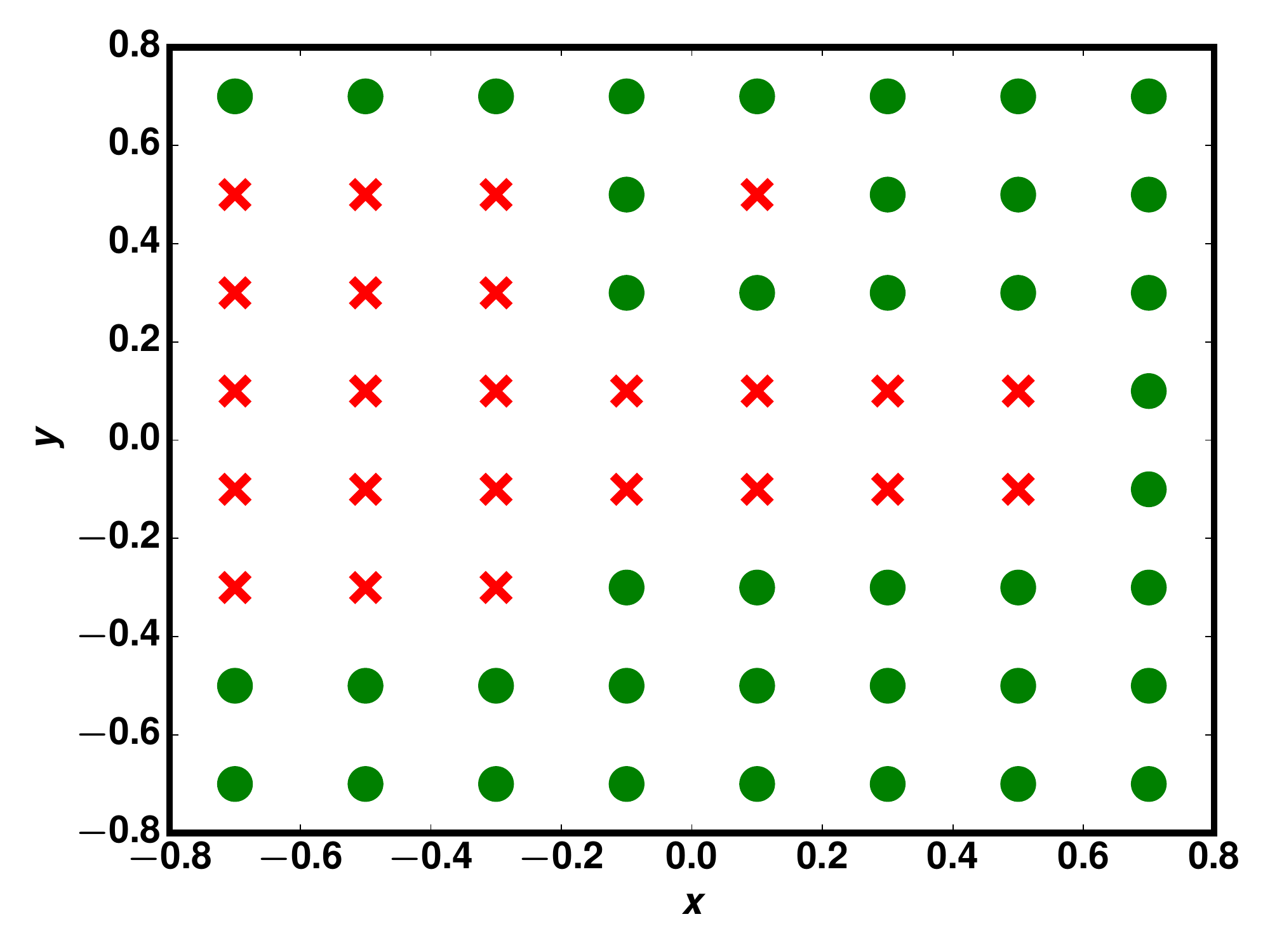}
		\caption{Coverage for Model 1}	
		\label{fish-fnn-coverage}
	\end{subfigure}	
	\begin{subfigure}{0.49\textwidth}
		\includegraphics[width=2.5in,height=2.2in]{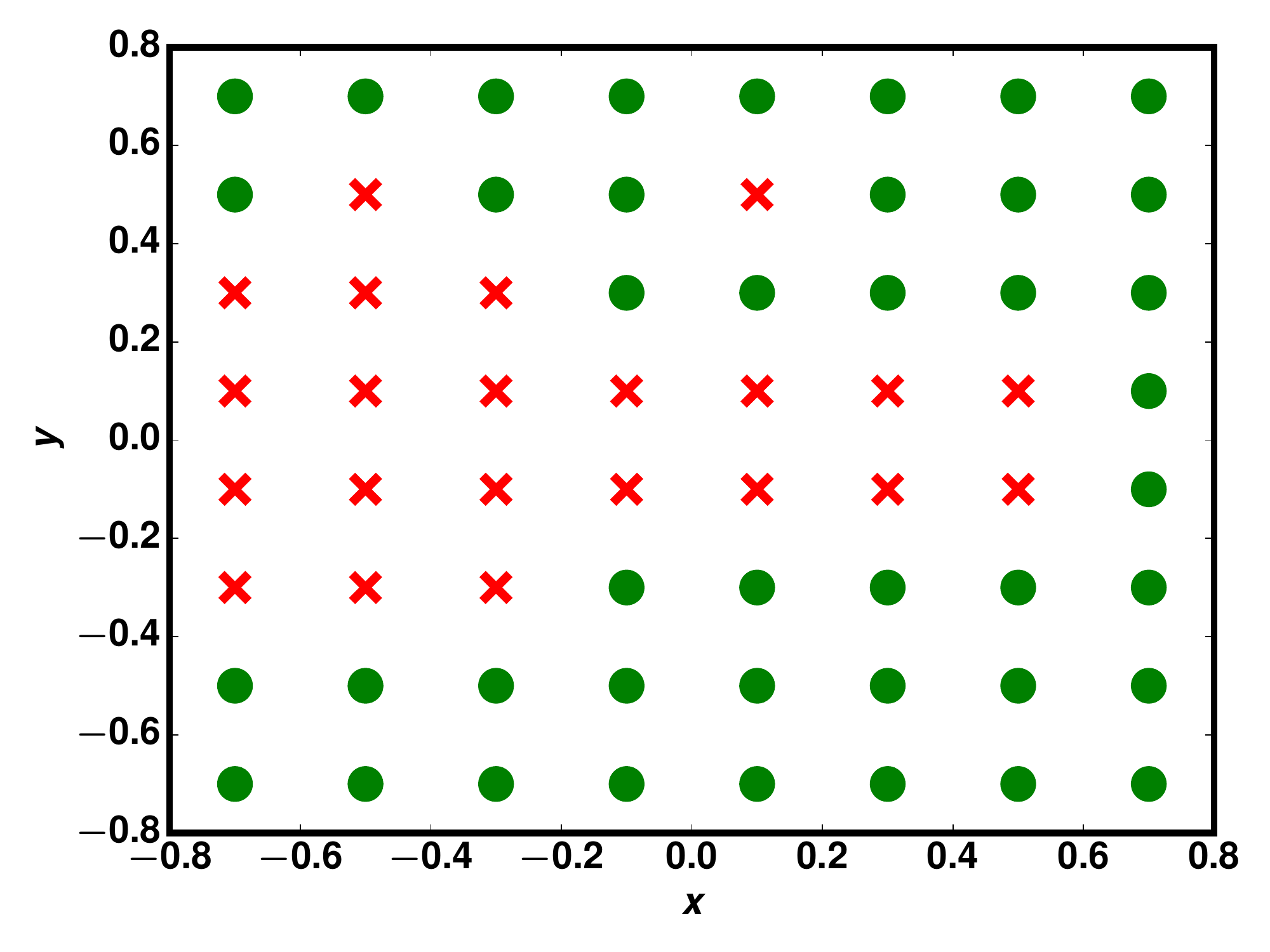}
		\caption{Coverage for Model 3}	
		\label{fish-rnn-coverage}
	\end{subfigure}	
	\caption{
		Coverages of the target location $(x^*, y^*, 0)$ for Models 1 and 3
		with two output neurons, after training.	 
		A green circle indicates a success while a red '$\times$' indicates
		a failure for the corresponding target location.
		(a) Model 1 with two output neurons.
		(b) Model 3 with two output neurons.  
		Both models have a similar range of coverage although Model 3 has a slightly larger region (65.63\%) than Model 1 does (62.50\%).
	}
	\label{fig:fish-rnn-coverage}
\end{figure}

Fig.~\ref{fig:fish-rnn-coverage} compares the coverages of target locations,
$(x^*, y^*, 0)$ for Model 1 and Model 3 controllers with two output neurons.
Coverage is defined as the set of target locations that the trained controller
could shepherd the fish to. Note that the initial orientation of the fish with
respect to the target location plays a big role in whether a target is
achivable, particularly when the target is very close.

\section{Conclusion} \label{conclusion}

We have proposed spiking neuron network controllers that are biologically
plausible and have applied them to learn the classical cart-pole control
problem as well as a fish locomotion control problem, to demonstrate their
efficacy.  The derivation of the synaptic update rule is general and can be
applied to any feedforward or recurrent network of spiking neurons.  The
experiments show that the proposed controllers have fairly large regions of
stability, and behave in a manner different from traditional PID controllers.
We have analyzed in detail multiple network architectures with different output
neuron settings: two output neurons with the same force magnitude (Model 1), 4
or more neurons with different force magnitude kernels (Model 2), and
recurrently connected neurons with multiple force kernels (Model 3).  From the
experiments, we deduce that more neurons with diverse force magnitudes can
learn larger coverage domains and are thus more flexible and robust.
Furthermore, we showed that the recurrent network controllers can produce
very sparse spike train outputs with firing rates as low as 0.9 Hz per neuron
signifying high energy efficiency. Finally, we have demonstrated that the
controller can learn even in a scenario where several control variables have to
be regulated synergistically to accomplish the control task.

Future directions that we wish to consider are: adding kernels for filtering
the process variable inputs, and more general control cost functions. The
former can readily be added to the current framework keeping the derivation
the same. The latter requires using more general error functions and analyzing
their impact on the resultant control behavior.

\subsection*{Acknowledgments}
The authors thank the Air Force Office of Scientific Research  (Grant FA9550-16-1-0135)  for their generous support of this research.




\bibliographystyle{apa-good}
\bibliography{spike_control}

%
%
%
%

\end{document}